\documentclass[aps,pre,groupedaddress,twocolumn]{revtex4-1}

\usepackage{graphicx}
\usepackage[utf8]{inputenc}
\usepackage[english]{babel}
\usepackage{amsmath,enumerate}

\usepackage{color}

\newcommand{\st}{\text{s}}

\newcommand{\out}{\text{out}}
\newcommand{\inn}{\text{inn}}

\begin{document}


\title{Hydrodynamics of granular particles on a line} 


\author{Andrea Baldassarri}
\author{Andrea Puglisi}
\affiliation{Istituto dei Sistemi Complessi - CNR and Dipartimento di Fisica, Universit\`a di Roma Sapienza, P.le Aldo Moro 2, 00185, Rome, Italy}
\author{Antonio Prados}
\affiliation{F\'isica Te\'orica, Universidad de Sevilla, Apartado de
  Correos 1065, E-41080 Sevilla, Spain}
\email[]{prados@us.es}


\date{\today}

\begin{abstract}
  We investigate a lattice model representing a granular gas in a thin
  channel.  We deduce the hydrodynamic description for the model from
  the microscopic dynamics in the large system limit, including the
  lowest finite-size corrections. The main prediction from
  hydrodynamics, when finite-size corrections are neglected, is the
  existence of a steady ``uniform longitudinal flow'' (ULF), with the
  granular temperature and the velocity gradient both uniform and 
  directly related. Extensive numerical simulations of the system show
  that such a state can be observed in the bulk of a finite-size
  system by attaching two thermostats with the same temperature at its
  boundaries.  The relation between the ULF state and the shocks
  appearing in the late stage of a cooling gas of inelastic hard rods
  is discussed.
\end{abstract}


\maketitle

\section{Introduction}\label{sec:intro}

Statistical physics offers a systematic
computational scheme for averages and fluctuations, suitable for
systems of particles at thermal equilibrium~\cite{huang}. A similar
scheme for systems far from  equilibrium is lacking and
represents an open challenge~\cite{m89}. An example of non-equilibrium
statistical system pervading many human activities is given by granular
fluids, an assembly of inelastic hard particles agitated by some
external driving mechanism~\cite{jaeger96b,poeschel,puglio15}.

The fluidised state of granular matter is an excellent testing ground
for kinetic theory~\cite{D01,goldhirsch03,brilliantov_kinetic_2004}. 
One of the most widely used model in granular kinetic theory is a
system composed of inelastic smooth hard particle
particles~\cite{brey_dissipative_1997,BDKS98,DB02}. 
In this context, the inelastic Boltzmann (or Enskog) equation has been
shown to be a powerful tool. However, the structure of the collision
term for inelastic hard particles makes the corresponding kinetic
equation a tough mathematical problem, for which exact solutions and
rigorous results are difficult to obtain.

In view of the above issue, models that simplify the collision term
have been proposed to make it easier their analytical
investigation, while preserving similar physics. This is the spirit of
the Maxwell-like collision model, either elastic \cite{ERNST19811} or
inelastic \cite{ben-naim_multiscaling_2000}, in which the collision
rate is assumed to be independent of the relative velocity.
Some experts claim that ``What harmonic oscillators
are for quantum mechanics, and dumb-bells for polymer physics, is what elastic and inelastic
Maxwell models are for kinetic theory''~\cite{EB02}.
Indeed, many rigorous results in the inelastic case have been derived,
both in the freely
cooling~\cite{ben-naim_multiscaling_2000,BMP02,ben-naim02,ernst_high-energy_2002,ernst_scaling_2002,bobylev_self-similar_2003,bobylev_proof_2003,bobylev_self-similar_2009,ilyin_exact_2016}
and the uniformly
heated~\cite{ben-naim_multiscaling_2000,carrillo_steady_2000,santos_exact_2003}
cases.
As pointed out in a review article by
Villani~\cite{villani_mathematics_2006}, the most relevant questions
in inelastic Maxwell models have been solved in homogeneous situations and,
rather than going for refinements, the current priority is to deal
with inhomogeneous states.

The nonconservation of energy has several, physically relevant,
implications. First, the distribution function is in general
non-Gaussian~\cite{van_noije_velocity_1998,SM09}, even for the
stationary states reached when some external mechanism injects energy
into the system. Second, and most importantly, kinetic theory
establishes a link between the microscopic and the macroscopic,
hydrodynamic, descriptions, by making it possible to derive the latter
from the former~\cite{BDKS98,D01,DB02}. Along this route, the
nonconservation of energy introduces another time scale that makes
more delicate the critical requirement on {\em separation of scales}
needed to accomplish
it~\cite{G99,K99,serero_hydrodynamics_2006}. However, there are many
situations, typically in dimension larger than one and for dilute
quasi-elastic systems, in which hydrodynamic equations fairly
reproduce real granular experiments qualitatively~\cite{argentina02}
or even quantitatively~\cite{lohse10,puglisi12}.

Here we aim at building and verifying the hydrodynamic equations for
an idealised one-dimensional (1d) lattice model~\cite{BMP02,bald2003}
with a Maxwell-like collision rule. The model simplifies the dynamics
of a granular fluid in a 1d channel with negligible density
fluctuations, a condition that reasonably holds in not too dilute
systems.  A strong connection between this idealised model and an
inelastic 1d gas of hard rods~\cite{ben-naim99} has been established
in previous studies~\cite{BMP02}. Specifically, there appears a
nontrivial correspondence between the velocity profiles of the lattice
model and the shock-like structures of the granular gas of hard rods
in the asymptotic cooling regime. The 1d lattice model has been
further studied in~\cite{ostojic04,dey_lattice_2011} and also
generalised to two dimensions~\cite{BMP02b}.

In spite of the above described strong connection between 1d granular
gases and the 1d Maxwell model, the hydrodynamic equations of the
latter have been neither written or analysed. Then, in the present
study, first we derive the hydrodynamic equations from the microscopic
dynamics. Afterwards, we focus on a solution of them: the steady
uniform longitudinal flow (ULF). We argue that such a state resembles
the shock profiles observed in the late cooling regime of many 1d
granular systems. Also, we incorporate finite-size corrections and
compare them with simulations of the microscopic dynamics.

Section~\ref{sec:overview} is devoted to a brief overview of previous
studies on granular models in $1d$ and their comparison with
hydrodynamic theories. Our model is introduced in
Section~\ref{sec_model}: in the same Section its hydrodynamic
equations are derived and the ULF solution is
discussed. Section~\ref{sec:numerics} presents the numerical results and their
comparison with theory. Finite size corrections and boundary layers
are discussed in Section~\ref{sec:finite-size-and-correlations}. Conclusions are drawn in
section~\ref{sec:conc}. Appendices illustrate the technical aspects of the
derivation of the hydrodynamics (Appendix A) and of the boundary layer
calculations (Appendix B), and give also some numerical results for
the local velocity distributions and spatial correlations (Appendix C)

\section{1d Granular hydrodynamics: a brief
  overview}\label{sec:overview}

Despite the
progress made during the last decades in the realm of granular kinetic
theory and hydrodynamics, the validity of a hydrodynamic description
in 1d systems is still under debate.  It should be stressed that 1d
systems essentially differ from higher dimensional setups in which the
gradients have a well-defined direction: particles cannot hop over
their nearest neighbours, which hinders the necessary ``mixing'' to
obtain a continuum description.

We briefly review some of the main results on 1d granular
hydrodynamics below. The 1d version of the granular Navier-Stokes
equations reads
\begin{subequations} \label{1dhd}
\begin{align} 
       D_t \rho = & -\rho \partial_x u,\\
  \rho D_t u= & - \partial_x P, \label{1dhdb}\\
  \rho D_t T = & - 2P \partial_x u-2 \partial_x q -\rho\zeta T,
\end{align}
\end{subequations}
where the density $\rho$, the mean velocity $u$, the granular
temperature $T$, the pressure $P$, the heat flux $q$ and the
cooling rate $\zeta$ depend on $(x,t)$. See, for instance,
\cite{astillero_unsteady_2012} for their microscopic definitions. As
usual, we have used the notation $D_t\equiv \partial_t + u \partial_x$
for the material derivative.

One of the first tests of 1d granular hydrodynamics was carried out
in~\cite{du95}. Therein, a gas of inelastic hard rods was confined by one or two thermostatted walls. Numerical
simulations revealed an incompatibility between Eqs.~\eqref{1dhd} and
the observed steady hydrodynamic profiles.  Later, the same system has been
demonstrated to lack a pure thermodynamic limit, i.e. infinite size at
constant restitution coefficient~\cite{PLMPV98,puglisi99}.

Subsequent studies of the 1d granular gas focused on the cooling
state, that is, with periodic boundary conditions and no external
driving. The 1d gas of hard rods develops strong
inhomogeneities, which include clusters and shocks, for long enough
times \cite{ben-naim99}. Extensive numerical simulations~\footnote{Therein, a regularisation of
  the collision was adopted: impacts at very small relative velocities
  were considered elastic to avoid inelastic
  collapse~\cite{luding98f}.} have revealed
that this 1d gas becomes then indistinguishable from a perfectly
``sticky''
gas~\cite{ben-naim99,shinde_violation_2007,shinde_equivalence_2009},
regardless of its actual inelasticity, a conjecture analysed also in
higher dimensions~\cite{nie02,trizac00} and for wet
particles~\cite{zaburdaev06}. Interestingly, the (inviscid) Burgers
equation $D_{t} u=\mu\partial_{x}^{2}
u$~\cite{nieuwstadt_selected_1995,su_kortewegvries_1969}, with $\mu\to
0$, describes the sticky
gas~\cite{shandarin_large-scale_1989,frachebourg99,frachebourg00}. Notwithstanding,
to the best of our knowledge, an analytical justification of the
Burgers equation in 1d granular gases has not been obtained yet. A
first step in this direction would be a neat derivation of 1d granular
hydrodynamics.

1d continuum equations can appear also in higher dimensions, when only
one direction develops a gradient, as in~\cite{brey01d} for a steady
case under gravity or in cooling systems with high aspect ratio without
gravity~\cite{meerson05b,meerson05,meerson07}.  In the latter case,
granular hydrodynamics predicts that the so-called {\em flow by
  inertia} $D_{t} u=0$ sets in during the highly inhomogeneous stage
of cooling. Therefore, there also appear shocks in $u(x,t)$, which
imply a singularity in $\rho(x,t)$. In addition, Molecular Dynamics
simulations of this channel system have made clear that such
singularities are approached following the hydrodynamic predictions,
that is, the flow by inertia scenario~\cite{meerson05}. Eventually,
when close-packing is almost reached, the flow by inertia picture
breaks down~\cite{meerson07,meerson08}.

\begin{figure}[b]
  \includegraphics[width=3.25in]{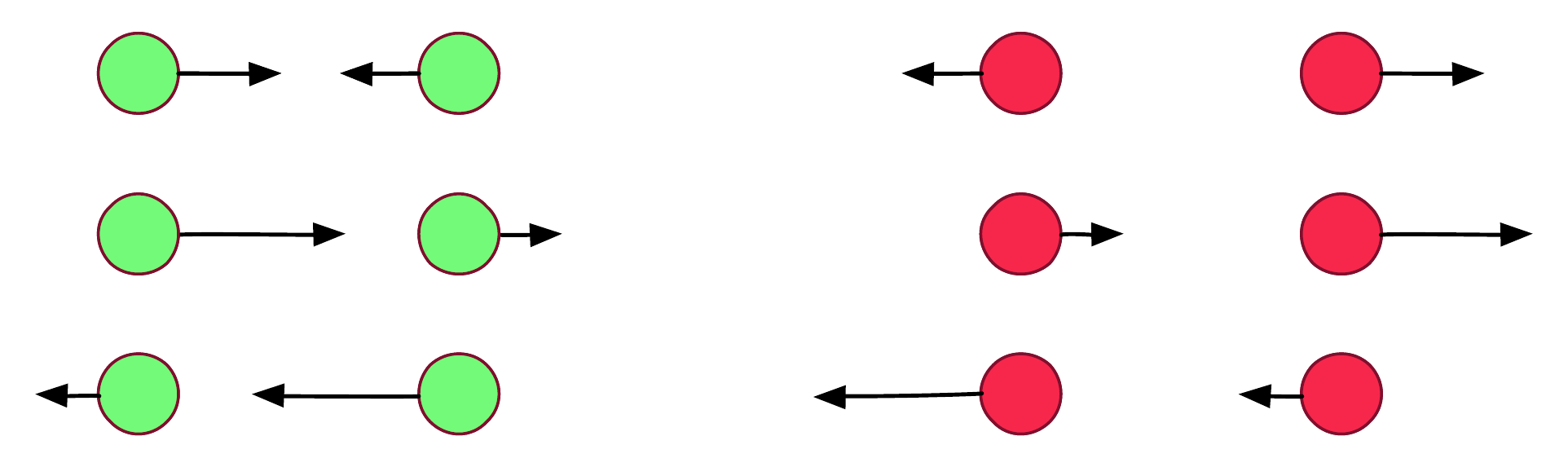}
  \caption{Allowed (green, left) and forbidden (red, right) collision,
    according to the kinematic constraint. Arrows show
    (precollisional) velocities of a pair of (candidate) colliding
    particles. Green pairs actually collides, red pairs do not
    collide. \label{fig:constr}\\ }
\end{figure}

\section{The model and its hydrodynamics}\label{sec_model}

Here, we study the 1d
lattice model first introduced in Refs.~\cite{BMP02,bald2003} to mimic
the evolution of a 1d granular gas, but neglecting density
fluctuations. The model is a 1d space-time discretised cellular
automaton, where we have one unit-mass particle at each site $l$,
$l=1,\ldots,N$. Particles are characterised by their (scalar)
``velocity'' $v_{l}$, which evolves according to a collisional
dynamics. Note that particles do not move but remain on the same site;
in this way density fluctuations are neglected.

At each discrete time step, a
random pair of adjacent sites is chosen and their corresponding
velocities, say $v$ and $\tilde{v}$, are updated according
to 
\begin{align}\label{collision}
          v' =& v+\frac{1+\alpha}{2} (\tilde{v}-v), &
  \tilde{v}' =&\tilde{v}-\frac{1+\alpha}{2} (\tilde{v}-v),
\end{align}
if $v$ and $\tilde{v}$ satisfy the so-called \textit{kinematic
  constraint}, that is, if the corresponding particles are ``approaching''
and not ``moving apart'', see Fig.~\ref{fig:constr}. In
Eq.~\eqref{collision}, $v'$ and $\tilde{v}'$ are the postcollisional
velocities, and $\alpha \in (0,1)$ is the restitution coefficient.
Momentum is conserved in collisions,
$v'+\tilde{v}'=v+\tilde{v}$, but energy is not:
$K'-K=\frac{1}{2}(\alpha^2-1)(\tilde{v}-v)^2<0$, with
$K=v^2+\tilde{v}^{2}$.  The choice of the (candidate) colliding pair
is uniformly distributed among all nearest neighbour pairs, as for
``pseudo-Maxwell''
molecules~\cite{ERNST19811,ben-naim_multiscaling_2000,villani_mathematics_2006,BMP02}.

The dropping of the kinematic constraint leads to a completely
different class of models {\em without kinematic
constraint}~\cite{prasad_high-energy_2013,
  prasad_driven_2014,
  LMPP15,MPLPP16,PMLPP16,prasad_driven_2016,
  prasad_velocity_2017,
  plata_global_2017,plata_kovacs-like_2017}.
This choice is justified by a different physical picture:
the on-site velocities on the lattice are thought as {\em transverse},
not longitudinal, velocities. This interpretation is confirmed by the
equivalence of the corresponding hydrodynamic equations with those for
the shear mode of inelastic gases in higher
dimension~\cite{LMPP15,MPLPP16}. Also, variants of the model without
kinematic constraint have been employed to analyse the dynamical
evolution of social and economic systems
\cite{BENNAIM2003190,BENNAIM200399,slanina_inelastically_2004,porfiri_decline_2007,bobylev_self-similar_2009,torok_opinions_2013,Iniguez2014}.

A physical comment on our adoption of non-moving particles on a
lattice is in order. This may be understood as if we were adopting a
Lagrangian coordinate in the actual 1d granular gas, by
characterising the particle positions by their index $i$ instead of
their real positions $x_{i}$. The corresponding
Lagrangian density profile, defined by the mass per unit of space
measured by $i$, would be uniform. Within this physical picture, the
partial time derivative $\partial_{t}$ in our model is expected to
play the role of the material derivative $D_{t}$ in the 1d granular
gas.


\subsection{Microscopic evolution of velocity and energy}\label{sec:micro-evol}

At discrete time $p$, the pair $\{y_p,y_{p}+1\}$ that may undergo a
collision is chosen at random and the velocity on site $l$ at time
$p+1$ is given by
\begin{equation} \label{microvel2}
v_{l,p+1}-v_{l,p}=-j_{l,p}+j_{l-1,p},
\end{equation}
where we have defined the microscopic momentum flux from site $l$ to
site $l+1$ at time $p$
\begin{equation} \label{micromc}
j_{l,p}=\delta_{y_p,l}\,\Theta(v_{l,p}-v_{l+1,p})\frac{1+\alpha}{2}(v_{l,p}-v_{l+1,p}).
\end{equation}
Therein, $\delta_{ij}$ is Kronecker's delta, which identifies the
colliding pair, and $\Theta(x)$ is Heaviside's step function, which
imposes the kinematic constraint. The random integer $y_{p}$ is
uniformly distributed in $[1,L]$, where $L$ is basically equal to $N$
but depends on the boundary conditions.  For a thermostatted system
$L=N+1$. Obviously, $j_{l,p}$ only differs from zero when
the pair $(l,l+1)$ actually collides. Note that no external volume
forces (like gravity, for instance) are being considered, but they could be incorporated by adding a term $f_{l,p}$ to the
rhs of Eq.~\eqref{microvel2}.  

The evolution of the  kinetic energy is obtained by squaring
Eq.~\eqref{microvel2}, which after some algebra yields
\begin{multline} \label{microen1}
  v_{l,p+1}^2-v_{l,p}^2 = \frac{\alpha^2-1}{4}\big[\delta_{y_p,l}\Theta(v_{l,p}-v_{l+1,p})(v_{l,p}-v_{l+1,p})^{2}\\
    \qquad+\delta_{y_p,l-1}\Theta(v_{l-1,p}-v_{l,p})(v_{l-1,p}-v_{l,p})^{2}\big]
    -J_{l,p}+J_{l-1,p}.
\end{multline}
The microscopic energy current is
$J_{l,p}=(v_{l,p}+v_{l+1,p})j_{l,p}$.  The sink terms in squared brackets
on the rhs of Eq.~\eqref{microen1}  stem from
the inelasticity of collisions and lead to a monotonic decrease of
the total energy. The last two terms are a discrete spatial derivative
and correspond to the energy flux  between neighbouring sites, which
is already present in the conservative case $\alpha=1$.

\subsection{Equations for the velocity and temperature fields in the continuum limit}\label{sec:continuum-eqs}

The average (over realisations) velocity field
$u_{l,p}=\langle v_{l,p} \rangle$ evolves according to
\begin{equation}\label{microvel3}
u_{l,p+1}-u_{l,p} = - \langle j_{l,p}-j_{l-1,p} \rangle.
\end{equation}
To compute the last average, we introduce a local equilibrium assumption
\begin{equation} \label{leq}
\mathcal{P}_2(v_l,v_{l+1}) \simeq \frac{1}{2\pi\sqrt{
    T_{l,p}T_{l+1,p}}}
e^{-\frac{(v_l-u_{l,p})^2}{2 T_{l,p}}-\frac{(v_{l+1}-u_{l+1,p})^2}{2 T_{l+1,p}}}.
\end{equation}
We also assume that $u_{l,p}$ and $T_{l,p}$ are smooth functions of
$l$ in the large system size limit $L\gg 1$, more specifically that
the discrete derivatives $u_{l+1,p} - u_{l,p}=O(L^{-1})$ and
$T_{l+1,p} - T_{l,p}=O(L^{-1})$, which gives
\begin{equation} \label{pre}
\langle \Theta(v_{l,p}-v_{l+1,p})(v_{l,p}-v_{l+1,p})\rangle \sim
\sqrt{T_{l,p}/\pi}. 
\end{equation}
This is a ``modified'' pressure, the usual ideal gas equation of state
$\rho T$ is replaced by $\sqrt{T}$ because (i) collisions do not occur
at a rate proportional to $\sqrt{T}$ but constant and (ii) density
$\rho$ is uniform, $\rho=1$. Therefore,
\begin{equation}
u_{l,p+1}-u_{l,p}\sim-\frac{1+\alpha}{2L}\left(\sqrt{T_{l,p}/\pi}-\sqrt{T_{l-1,p}/\pi}\right).
\end{equation}

Now we introduce a continuum limit by defining  spatial and
temporal variables as
\begin{equation}\label{eq:cont-var}
  x=\epsilon\,l, \qquad t=\epsilon^{2}\,p, \qquad \epsilon=L^{-1}\ll 1,
\end{equation}
to obtain 
\begin{equation} \label{macrovel1}
\partial_t u(x,t) = -\frac{1+\alpha}{2}\partial_x \sqrt{T(x,t)/\pi}+O(\epsilon).
\end{equation}

We now discuss the average of Eq.~\eqref{microen1} for the kinetic
energy, which contains both dissipative and transport terms.  Within the local
equilibrium approximation, the average of the dissipative or cooling
term is
\begin{equation}
  \frac{\alpha^2-1}{4L}(T_{l,p}+T_{l-1,p})=\frac{(\alpha^2-1)\epsilon}{2}\,
  T_{l,p}+O\left((\alpha^2-1)\epsilon^{2}\right),
\end{equation}
which in the continuum limit reads
\begin{equation} \label{discont}
  \frac{(\alpha^2-1)\epsilon}{2}T(x,t)+O\left((\alpha^2-1)\epsilon^{2}
  \right).
\end{equation}
This linear cooling, instead of the typical $T^{3/2}$ behaviour, also
stems from the constant collision rate.

The average of the energy current, again under local equilibrium, reads
$  \langle J_{l,p} \rangle = (1+\alpha)\,\epsilon\,  u_{l,p}\sqrt{T_{l,p}/\pi} + O(\epsilon^{2})$,
so that the average of the last two terms in Eq.~\eqref{microen1} goes
in the continuum limit to
\begin{equation} \label{transcont}
  -(1+\alpha)\,\epsilon^{2}\,\partial_x\left[u(x,t)\sqrt{T(x,t)/\pi}\right]+
  O(\epsilon^{3}).
\end{equation}
Comparing Eqs.~\eqref{discont} and~\eqref{transcont}, we see that they
are of the same order only when $\alpha^2-1=O(\epsilon)$:
this choice makes them of order $\epsilon^{2}$, consistently with the
scaling of the finite-time difference on the lhs of
Eq.~\eqref{microen1}, $\Delta t=\epsilon^{2}$. Therefore, we introduce a ``macroscopic inelasticity''
\begin{equation}\label{eq:nu}
  \nu=L(1-\alpha^2)/2 \ge 0.
\end{equation}
Taking into account that $\langle
v^{2}\rangle=u^{2}+T$, we get
\begin{equation} \label{macrotemp1}
\partial_t T(x,t)=-\nu T(x,t)-2\sqrt{T(x,t)/\pi}\,\partial_xu(x,t) +O(\epsilon).
\end{equation}
where we have also made use of $1+\alpha \sim 2$ for $L\gg 1$ with
constant $\nu$. Moreover, Eq.~\eqref{macrovel1}
becomes
\begin{equation} \label{macrovel2}
\partial_t u(x,t) = -\partial_x \sqrt{T(x,t)/\pi}+O(\epsilon).
\end{equation}

Equations \eqref{macrovel2} and \eqref{macrotemp1} are the
hydrodynamic equations of the model to the lowest order. They coincide
with the hydrodynamic Eqs.~\eqref{1dhd} after identifying:
$D_t \to \partial_t$ (Lagrangian coordinate), $\rho \to 1$,
$P \to \sqrt{T/\pi}$ (Eq.~\eqref{pre}), $\zeta \to \nu$ and
$\partial_x q=0$.  The absence of dissipative transport (viscous
stress and heat flow) makes these equations close relatives of the
so-called Ideal Granular
Hydrodynamics~\cite{meerson07,meerson08,rozanova}.

\subsection{Steady uniform longitudinal flow}\label{sec:ULF}

In the large system size limit, the stationary solution of
Eqs.~\eqref{macrovel2} and~\eqref{macrotemp1} can be obtained. It
suffices to ask that the velocity flow in the middle of the system is
$0$ ~\footnote{By reason of symmetry or, alternatively, by imposing
  that total momentum vanishes, $\int_{0}^{1}dx\,u(x,t)=0$.}  to get
the following steady ULF profiles
\begin{equation}\label{statprof}
T_{\st}(x)=T_0, \qquad
u_{\st}(x)=\frac{\nu}{2} \sqrt{\pi T_0}\left(\frac{1}{2} - x\right),
\end{equation}
in which the constant $T_{0}$  remains undetermined at this
level of description \footnote{From a mathematical point of view, this
  property stems from the fact that the hydrodynamic equations contain
  only first-order spatial derivatives, which make it impossible to
  fit all the boundary conditions at $x=0,L$
  \cite{bender_advanced_1999}.}. The steady ULF in
Eq.~\eqref{statprof} does not require the presence of thermostats at
the boundaries, it is {\em self-sustained}.  Nevertheless, as
demonstrated numerically below, one can recover the above steady ULF
profile at the system bulk in a finite-size system by attaching two
identical thermostats at the boundaries. Also, note that this ULF, at
difference with the one discussed in
Ref.~\cite{astillero_unsteady_2012}, is incompressible because no mass
flow is allowed in our lattice system by definition.

Quite strikingly, regions with a linear velocity profile with negative
slope are also observed in the Lagrangian coordinate during the
formation of shocks, in numerical simulations of the inelastic 1d gas
of hard rods~\cite{ben-naim99,BMP02},
the inelastic hard disk gas in a channel~\cite{meerson05} and the
lattice model~\cite{BMP02}, all in the cooling regime at long
times. This suggests an interesting connection between the ULF and the
shocks characterising granular cooling in 1d (or quasi-1d)
systems. 

\section{Numerical results}\label{sec:numerics}

Numerical simulations of the model have
been carried out and compared with the previous theory. A system with
$N$ particles is initialised with uncorrelated, normally distributed
random velocities.  We introduce thermostats at its boundaries, at
sites $0$ and $N+1$, to make the system reach a steady state. Then, at
each time step $p$, the uniformly distributed random integer $y_{p}$
choosing the candidate colliding pair $\{y_p,y_p+1\}$ is drawn between
$0$ and $N$, that is, $L=N+1$. The velocities for the peripheral sites
of indexes $0,L$ are randomly and independently drawn from a normal
distribution with zero average and unit variance~\footnote{The
  temperature of the thermostat only sets the scale of energy and
  therefore we are not losing any generality.}. Due to the kinematic
constraint, the collision described by Eq.~\eqref{collision} only
takes place if $v_{y_p}-v_{y_p+1}>0$. Otherwise, velocities remain
unchanged. A large number of independent long runs has been performed
to get average stationary profiles of velocity and energy. Typically,
our data correspond to $10^5$ runs of more than $10^2 L^2$ steps each,
starting from independent normal-distributed velocities.

\begin{figure}
  \includegraphics[clip=true,width=3.25in]{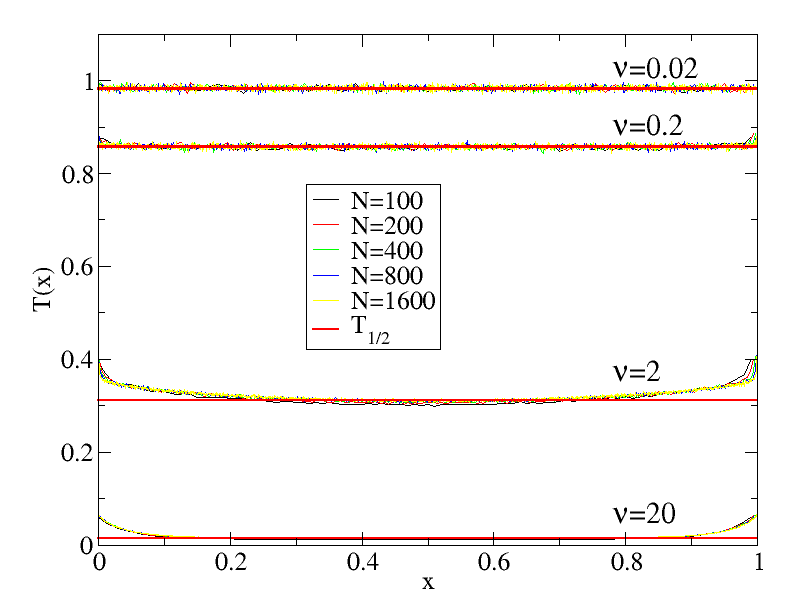}
  \caption{\label{fig:temp-prof} Average profile for the steady
    temperature, for different values of the macroscopic inelasticity
    $\nu$. The solid (red) lines stand for the estimated value of the
    temperature $T_{1/2}$ in the system bulk. 
  }
\end{figure}

Our numerical results for the average temperature profiles are shown
in Fig.~\ref{fig:temp-prof} for different values of the macroscopic
inelasticity $\nu$. From them, we measure the temperature of the
largest system at mid position, denoting it by $T_{1/2}\equiv
T(x=1/2)$. Later, we compare $T_{1/2}$ with the bulk temperature $T_0$ of Eq.~\eqref{statprof}, which predicts a constant temperature
profile, in quite good agreement with our observed numerical profiles.

\begin{figure}
\includegraphics[clip=true,width=3.25in]{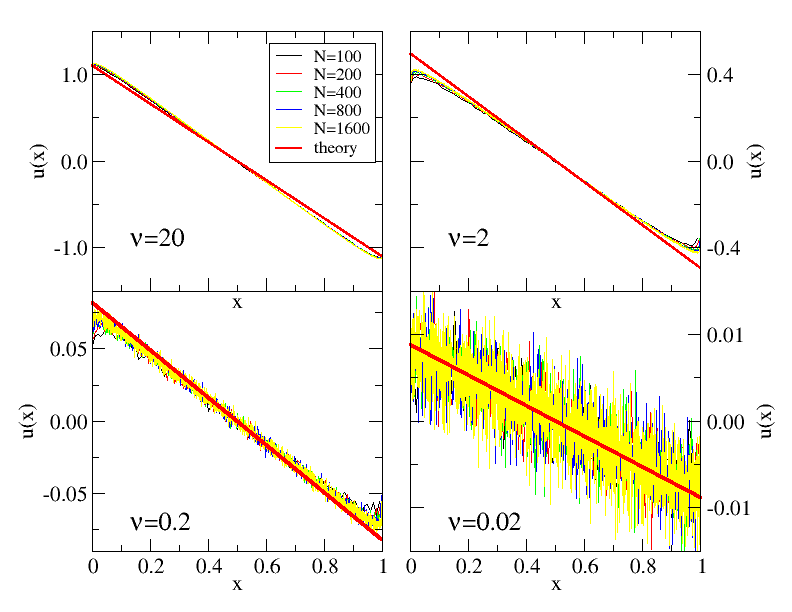}
\caption{\label{fig:veloc-prof} Average profile for the stationary
  velocity, for the same values of $\nu$ considered in
  Fig.~\ref{fig:temp-prof}. The solid (red) lines stand for the
  theoretical expression for $u_{\st}(x)$ in Eq.~\eqref{statprof},
  with $T_{0}$ given by the numerical estimate $T_{1/2}$ for the bulk
  temperature in Fig.~\ref{fig:temp-prof}. }
\end{figure}

Numerical results for the average velocity profiles are reported in
Fig.~\ref{fig:veloc-prof}. Again, the comparison with the theory is
satisfactory: in the bulk, the profiles are linear with a slope almost
perfectly matching the one given by the value of $T_{1/2}$ in
Fig.~\ref{fig:temp-prof}.  Both for the temperature and the average
velocity, deviations from the theoretical ULF profiles are apparent
near the boundaries, especially for $\nu=2$. See
below for a closer look at this issue.

\section{Finite size corrections and boundary layers}\label{sec:finite-size-and-correlations}

In Appendix \ref{app:A}, we show how to incorporate finite size
corrections. Still, we do so assuming local equilibrium, in order to
estimate in the simplest way the effect of these higher order
terms. Specifically, these $O(\epsilon)$ corrections introduce
second-order spatial derivatives in the evolution equations, and these
viscous terms make it possible to accommodate all the boundary
conditions. The considered thermostats impose that
\begin{equation}\label{eq:bc}
  u(0,t)=u(1,t)=0, \qquad T(0,t)=T(1,t)=1.
\end{equation}

The smallness of the viscous terms bring to bear two boundary layers
close to the thermostats at $x=0,1$, see Appendix \ref{app:B} for
details. Therein, we determine the
unknown bulk temperature $T_{0}$ by asymptotic matching
\cite{bender_advanced_1999}, which yields
\begin{equation} \label{bulkt}
\varphi(T_{0})=\frac{\sqrt{\pi}}{4}\nu,\; \textrm{with}\;\; \varphi(T_{0}) \equiv \sqrt{\frac{1}{3} + \frac{2}{3}T_{0}^{-3/2}-T_{0}^{-1}}.
\end{equation}
The function $\varphi(T_{0})$ decreases monotonically
from infinity to zero as $T_{0}$ is varied from zero to
unity. Therefore, Eq.~\eqref{bulkt} tells us that $T_{0}$ is a
monotonically decreasing function of the macroscopic inelasticity
$\nu$, with
\begin{equation}\label{eq:bulkt-lims}
  \lim_{\nu\to 0} T_{0}=1, \qquad \lim_{\nu\to\infty} T_{0}=0.
\end{equation}
These two limit results are expected on a physical basis: in the
elastic limit $\nu\to 0$, the system should be in equilibrium at the
temperature of the heat baths whereas in the strongly dissipative
limit $\nu\to\infty$ all the energy is dissipated in the boundary
layers before reaching the bulk, which is then at zero
temperature~\footnote{The width of the boundary layers typically
  scales as $\nu^{-1/2}$, as shown for instance in
  Ref.~\cite{prados_nonlinear_2012} for the dissipative version of the
  Kipnis-Marchioro-Presutti model.}.

\begin{figure}
\includegraphics[clip=true,width=3.25in]{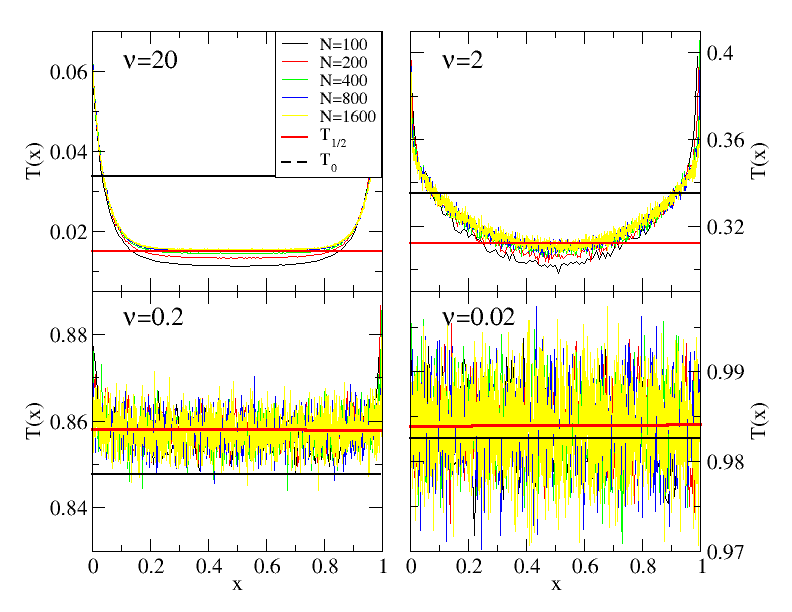}
\caption{\label{t_prof2} Average temperature profiles $T(x)$ in the
  stationary state for different values of $\nu$ (blow up of data in
  Fig.~\ref{fig:temp-prof}). Red lines stand for the numerical
  estimate $T_{1/2}$ for the bulk temperature, see
  Fig.~\ref{fig:temp-prof}, whereas black lines give our analytical
  prediction $T_{0}$, as given by Eq.~\eqref{bulkt}. }
\end{figure}

A more precise inspection of the temperature profiles is shown in
Fig.~\ref{t_prof2}, aimed to emphasise the deviation from the uniform
bulk profile in Eq.~\eqref{statprof}. Interestingly, the estimate for
the bulk temperature $T_{0}$ stemming from our finite size analysis,
as given by Eq.~\eqref{bulkt}, is quite close to our numerical
estimation $T_{1/2}$.  Most importantly, the success of our theory is
good for very different values of $\nu$, both small and large.  The
absolute error $|T_{0}-T_{1/2}|$ increases very slowly with $\nu$,
ranging from $0.01$ for $\nu=0.2$ to $0.02$ for $\nu=20$. The relative
error is also reasonable, remaining under ten per cent for
$\nu \lesssim 2$.

In Appendix \ref{app:C}, we look into possible sources of discrepancy
between simulations and theory. We focus on violations of local
equilibrium, Eq.~\eqref{leq}: both non-Gaussian local velocity
distributions and nearest-neighbour correlations. Our preliminary
conclusion is that the former are more important than the
latter, see Appendix \ref{app:C}.

\section{Concluding remarks}\label{sec:conc}

We have analysed a 1d lattice model, which by construction do not
contain density fluctuations.  Our derivation of hydrodynamics can be
thought as if carried out in an off-lattice system but using a
Lagrangian coordinate at the microscopic level.  The resulting
Eqs.~\eqref{macrovel2} and~\eqref{macrotemp1}, in which finite-size
corrections have been neglected, tell us that a self-sustained ULF
steady state appears in the bulk. Therein, the granular temperature is
uniform and the average velocity has a linear profile, whose slope is
directly related to the temperature. In order to observe such a steady
state in a finite-size system, thermostats at the boundaries must be
introduced, as shown in our numerical simulations which fairly compare
with our theory. The dependence of the bulk temperature on the
temperature of the thermostats is only obtained after incorporating
finite-size corrections to the theory.

Our work demonstrates that 1d granular hydrodynamic equations, whose
validity is still under debate, can be derived by means of a proper
continuum limit that keeps a finite macroscopic inelasticity, but in
which the microscopic dynamics is quasielastic.  Our main assumption
is the local equilibrium approximation, which has been shown to be
valid in other simple models for a wide range of the system parameters
\cite{prados_nonlinear_2012,LMPP15,MPLPP16}. How to improve upon the
current results by going beyond local equilibrium is an open
perspective for future work.

An emerging interesting conjecture is whether the system in the
cooling regime is still well described by our hydrodynamic
equations. In fact, the ULF discussed here seems to be similar to the
linear velocity profiles in shock regions of 1d cooling granular
gases, specifically for either hard rods on a line~\cite{ben-naim99}
or hard disks in a channel geometry~\cite{meerson05}. A verification
of this conjecture, together with a route to connect our theory with
the Burgers or flow-by-inertia equations, is currently under
investigation.

\begin{acknowledgments}
We acknowledge useful discussions with Dario Villamaina. A.P. acknowledges the support of Universidad de Sevilla’s VI Plan Propio de Investigación through Grant PP2018/494.
\end{acknowledgments}

\appendix

\section{Hydrodynamic equations with finite size corrections}\label{app:A}

Here we derive the hydrodynamic equations, incorporating into them
finite size, $O(L^{-1})$, corrections. First, we analyse the evolution
equation for the velocity: our starting point is
Eq.~(\ref{microvel3}), in which the average
momentum current at site $l$ is exactly given as
\begin{equation}\label{eq:av-veloc-current}
  \langle
j_{l,p}\rangle=\frac{(1+\alpha)\epsilon}{2}\langle\Theta(v_{l,p}-v_{l+1,p})(v_{l,p}-v_{l+1,p})\rangle,
  \;\; \epsilon=L^{-1},
\end{equation}
since $y_{p}$ is an independent, uniformly distributed,
stochastic integer choosing the specific pair that collides. We recall
that $L$ is the total number of possible colliding pairs, whose
relation to the number of sites $N$ depends on the boundary
conditions.  The average on the rhs of the above equation is done over
all the velocities $\{v_{l,p},v_{l+1,p}\}$, by assuming the local
equilibrium approximation written in Eq.~(\ref{leq}) of the main
text. The result is
\begin{widetext}
\begin{equation}\label{av-veloc-curr-LE}
  \langle\Theta(v_{l,p}-v_{l+1,p})(v_{l,p}-v_{l+1,p})\rangle=
  \frac{u_{l,p}-u_{l+1,p}}{2}
  \left[\text{erf}\left(\frac{u_{l,p}-u_{l+1,p}}{\sqrt{2
        (T_{l+1,p}+T_{l,p})}}\right)+1\right]+ \exp \left[-\frac{(u_{l,p}-u_{l+1,p})^2}{2 (T_{l+1,p}+T_{l,p})}\right] \sqrt{\frac{T_{l+1,p}+T_{l,p}}{2\pi}},
\end{equation}
\end{widetext}
where $\text{erf}(z)$ is the error function defined by
\begin{equation}\label{eq:error-func}
  \text{erf(z)}=\frac{2}{\sqrt{\pi}}\int_{0}^{z} dt \, e^{-t^{2}}.
\end{equation}

Now we go to a continuum limit in space by assuming that both
$u_{l,p}$ and $T_{l,p}$ vary smoothly with the site index $l$. Then,
we define a continuous spatial variable $x=\epsilon l$, with
$\epsilon=L^{-1}\ll 1$ and make the mapping $u_{l,p}\to u(x;p)$,
$T_{l,p}\to T(x;p)$ and, consistently,
$u_{l\pm 1,p}\to u(x\pm\epsilon;p)$,
$T_{l\pm 1,p}\to T(x\pm \epsilon;p)$. Of course, this is also done for
the current, $j_{l,p}\to j(x;p)$ and
$j_{l\pm 1;p}\to j(x\pm\epsilon;p)$. In addition, we expand all the
functions evaluated at $x\pm \epsilon$ in powers of the small
parameter $\epsilon$, which transforms the exact momentum balance
Eq.~(\ref{microvel3}) into
\begin{align}
 & u(x;p+1)-u(x;p)=-\frac{(1+\alpha)\epsilon^{2}}{2} \nonumber \\ 
 & \qquad\quad  \times \left[\partial_{x}\sqrt{\frac{T(x;p)}{\pi}}-\frac{\epsilon}{2}  \partial_{x}^{2}u(x;p)+O\left(\epsilon^{2}\right)\right].
\end{align}
It is clearly seen that the evolution of $u(x;p)$, being proportional
to $\epsilon^{2}$, is very slow in discrete time. This suggests the
introduction of a continuous time scale $t=p \Delta t$, with
$\Delta t=\epsilon^{2}$~\footnote{In other lattice models, the typical
  scaling has been found to be $t\propto \epsilon^3$ as in
  Refs.~\cite{prados_nonlinear_2012,lasanta_fluctuating_2015,manacorda_lattice_2016}.}.
Over this scale,
$u(x;p+1)-u(x;p)=\Delta t \, \partial_t
u(x,t)+O\left(\epsilon^{4}\right)$ and we thus finally have
\begin{equation}\label{eq:u-evol}
    \partial_{t}u(x,t)=\frac{1+\alpha}{2}\left[
      -\partial_{x}\sqrt{\frac{T(x,t)}{\pi}}+\frac{\epsilon}{2}
      \partial_{x}^{2}u(x,t)+O\left(\epsilon^{2}\right) \right].
\end{equation}

Next, we repeat the above procedure for the balance of energy
equation. The corresponding  expressions are much lengthier than those
for the average velocity above and thus we only give the final
expressions for the averages of the dissipative and flux terms on the
rhs of Eq.~(\ref{microen1}). For the dissipative term, we have
\begin{widetext}
\begin{align}
\frac{(\alpha^2-1)\epsilon}{4}\langle\Theta(v_{l,p}-v_{l+1,p})(v_{l,p}-v_{l+1,p})^{2}+\Theta(&v_{l-1,p}-v_{l,p})(v_{l-1,p}-v_{l,p})^{2}\rangle
                                                                                       \nonumber
  \\
  &=\frac{(\alpha^{2}-1)\epsilon}{2}\left[
  T(x;p)-2\epsilon\,  \sqrt{\frac{T(x;p)}{\pi}}
  \partial_{x}u(x;p)\right]+O((\alpha^{2}-1)\epsilon^{3}),
\end{align}
again within the local equilibrium approximation. For the flux terms,
\begin{equation}
  \langle J_{l,p}-J_{l-1,p}\rangle=(1+\alpha)\epsilon^{2}\left\{ \partial_{x}\left[ u(x;p) \sqrt{\frac{T(x;p)}{\pi}}\right] -\frac{1}{4}
   \epsilon \left\{\partial_{x}^{2}T(x;p)+2\, \partial_{x}\left[u(x;p)
     \partial_{x}u(x;p)\right]\right\}+O\left(\epsilon^{2}\right)\right\}.
\end{equation}
The evolution equation for the energy $e$ is then
\begin{eqnarray}
  \langle e(x;p+1) -e(x;p)\rangle&=& \frac{(\alpha^{2}-1)\epsilon}{2}\left[
                                     T(x;p)-2\epsilon\, \sqrt{\frac{T(x;p)}{\pi}} \partial_{x}u(x;p)
                                     \right]
                                     \nonumber \\
 && -(1+\alpha)\left\{ \epsilon^{2 }\partial_{x}\left[ u(x;p) \sqrt{\frac{T(x;p)}{\pi}}\right] -\frac{1}{4}
   \epsilon^{3} \left\{\partial_{x}^{2}T(x;p)+2 \partial_{x}\left[u(x;p)
    \partial_{x}u(x;p)\right]\right\}\right\} \nonumber \\
 &&  +O((\alpha^{2}-1)\epsilon^{3}) +O(\epsilon^{4}).
\end{eqnarray}
Going again to the continuous time variable,
\begin{eqnarray}
  \partial_{t}e(x,t)+O(\epsilon^{2})&=&\frac{\alpha^{2}-1}{2\epsilon}\left[
                                       T(x,t)-2\epsilon\,                                         \sqrt{\frac{T(x,t)}{\pi}}\partial_{x}u(x,t)
                                        \right]-(1+\alpha)\partial_{x}\left[ u(x,t) \sqrt{\frac{T(x,t)}{\pi}}\right]
                                        \nonumber \\
                                    &&  +\frac{1+\alpha}{4}
                                       \epsilon \left\{\partial_{x}^{2}T(x;p)+2 \partial_{x}\left[u(x;p)
                                       \partial_{x}u(x;p)\right]\right\}
                                        +O(\epsilon^{2})+O((\alpha^{2}-1)\epsilon).
\end{eqnarray}
\end{widetext}
In order to have a consistent limit over the continuous time $t$,
$\alpha^{2}-1$ must be of the order of $\epsilon$. Hence, we introduce
the macroscopic inelasticity $\nu$, as defined in Eq.~\eqref{eq:nu},
which is assumed to be of order of unity. This implicitly assumes that
the underlying microscopic dynamics is quasi-elastic, since
$1-\alpha\sim\nu\epsilon=\nu/L \ll 1$ \footnote{Again, this scaling
  for the macroscopic inelasticity is different from the one found in
  other models, as a consequence of the different scaling of the
  continuous time variable. Notwithstanding, the underlying
  microscopic dynamics is quasi-elastic in all cases.}. With this
definition,
\begin{widetext}
\begin{align}
  \partial_{t}e(x,t)=&-\nu
  T(x,t)-(2-\nu\epsilon)\partial_{x}\left[u(x,t)\sqrt{\frac{T(x,t)}{\pi}}\right]+\epsilon
  \left\{2\nu\sqrt{\frac{T(x,t)}{\pi}}\partial_{x}u(x,t)+\frac{1}{2}\partial_{x}^{2}T(x,t)+\partial_{x}\left[u(x,t)\partial_{x}u(x,t)\right]\right\}
                       \nonumber \\
  & +O(\epsilon^{2}).
\end{align}

Finally, the hydrodynamic equations for the average velocity and the
temperature up to order $\epsilon=L^{-1}$ are obtained by taking into
account the definition of the macroscopic inelasticity \eqref{eq:nu}
and the identity $e(x,t)=u^{2}(x,t)+T(x,t)$, which yields
\begin{subequations}\label{eq:hydrodynamics}
\begin{equation}\label{eq:u-evol-bis}
   \partial_{t}u(x,t)=
      -\partial_{x}\sqrt{\frac{T(x,t)}{\pi}}+\frac{\epsilon}{2} 
      \left[ \nu\, \partial_{x}\sqrt{\frac{T(x,t)}{\pi}}+ \partial_{x}^{2}u(x,t) \right]+O\left(\epsilon^{2}\right) ,
\end{equation}
\begin{equation}\label{eq:temp-evol}
  \partial_{t}T(x,t)=-\nu
  T(x,t)-2\sqrt{\frac{T(x,t)}{\pi}}\partial_{x}u(x,t)+\epsilon
  \left\{3\nu\sqrt{\frac{T(x,t)}{\pi}}\partial_{x}u(x,t)+\frac{1}{2}\partial_{x}^{2}T(x,t)+\left[\partial_{x}u(x,t)\right]^{2}\right\}+O(\epsilon^{2}).
\end{equation}
\end{subequations}
\end{widetext}

Note that the parameter $\epsilon$ can be understood as the ratio of a
microscopic length ($1$ lattice site) to a macroscopic length (the
total length $L$). In this way, the condition $\epsilon\ll 1$ can be
conceived as the usual small Knudsen number condition for the validity
of a continuum description. A more detailed look at this issue can be
found in the following section.

\section{Boundary layer calculations}\label{app:B}

Now we solve the hydrodynamic equations to the lowest order, but
incorporating the boundary layers close to the system edges. Thus, we
consider Eqs.~\eqref{eq:hydrodynamics} with the
boundary conditions in the half interval $[0,1/2]$
\begin{subequations}\label{eq:evol-bc}
\begin{align}
  & u(0,t)=0,  && T(0,t)=1,  \label{eq:evol-bc-a} \\
  & u(1/2,t)=0,   && \left.\partial_{x}T(x,t)\right|_{x=1/2}=0.
\end{align}
\end{subequations}
The boundary conditions at $x=1/2$ stem from $u(x,t)$ and $T(x,t)$
being an odd and an even function with respect to $x=1/2$, respectively. 

We focus on stationary solutions of the hydrodynamic equations. Then,
we can simplify the notation by introducing
$\prime\equiv\partial_{x}$. Using the terminology in Ref.~\cite{bender_advanced_1999}, the
``outer'' solutions $\{u_{\out}(x),T_{\out}(x)\}$ satisfy the equations
\begin{equation}\label{eq:outer-eqs}
  \left( \sqrt{\frac{T_{\out}}{\pi}}\right)^{\prime}=0, \quad -\nu
  T_{\out}-2\sqrt{\frac{T_{\out}}{\pi}} u_{\out}^{\prime}=0.
\end{equation}
These equations are first-order in space and thus we can only impose
the boundary conditions at one of the endpoints of the half interval
$[0,1/2]$, specifically those at the system centre $x=1/2$. Then,
\begin{equation}\label{eq:outer-sols}
  T_{\out}=T_{0}, \quad u_{\out}(x)=-\frac{\nu}{2}\sqrt{\pi\, T_{0}} \left(x-\frac{1}{2}\right),
\end{equation}
where $T_{0}$ remains undetermined.

The explicit form of the outer (bulk) solutions allows us to have a
more precise look at the necessary condition to have a meaningful
continuum description. Making use of Eq.~\eqref{eq:outer-sols}, we can
readily identify a macroscopic length $\xi=(\nu\sqrt{\pi}{2})^{-1}$
(the inverse of the factor in front of the expression for
$u_{\out}(x)$, apart from $\sqrt{T_{0}}$ that carries with it the
velocity dimensions). The microscopic length scale in the continuum is
$\epsilon=L^{-1}$, corresponding to one lattice site, and then the
sought condition would be
\begin{equation}
  \frac{\epsilon}{\xi}=\frac{\nu\sqrt{\pi}}{2L}=
  \frac{(1-\alpha^{2})\sqrt{\pi}}{4}\ll 1,
\end{equation}
where we have again used the definition of $\nu$, Eq.~\eqref{eq:nu}. This
is nothing but a condition of quasielasticity for the microscopic
dynamics, a restriction that is already known for other lattice models
\cite{prados_nonlinear_2012,lasanta_fluctuating_2015,MPLPP16}. Several
comments are in order. First, note that this condition has already
been imposed when we have derived the hydrodynamic equations,
specifically when we have assumed that the macroscopic inelasticity
$\nu$ is of the order of unity. Second, this is physically reasonable:
the magnitude of the gradients in granular systems is often typically
controlled by the inelasticity, not by the imposed boundary
conditions. Therefore, the microscopic inelasticity $1-\alpha$ must be
small in lattice models to ensure that the gradients are not so large
that the lattice is unavoidable ``seen'' and a continuum description
is not possible. Third, despite the quasielasticity of the microscopic
dynamics, the observed macroscopic behaviour does not correspond to a
quasielastic granular fluid, because the macroscopic inelasticity $\nu$
(also called the cooling rate) is not small.

A different distinguished limit is obtained within the boundary layer,
with the scaling
\begin{equation}\label{eq:bl-scaling}
  x=\epsilon X, \qquad \frac{d}{dx}=\epsilon^{-1}\frac{d}{dX}.
\end{equation}
It is $X$ that the ``inner'' solution within the boundary layer, which
we denote by $\{u_{\inn}(X),T_{\inn}(X)\}$, depends on. The dominant
terms in the balance equations become
\begin{subequations}\label{eq:bl-eqs}
  \begin{equation}\label{eq:bl-eq-1}   -\frac{d}{dX}\left(\sqrt{\frac{T_{\inn}}{\pi}}\right)+\frac{1}{2}\frac{d^{2}}{dX^{2}}u_{\inn}=0,
  \end{equation}
  \begin{equation}\label{eq:bl-eq-2}   -2\sqrt{\frac{T_{\inn}}{\pi}}\frac{d}{dX}u_{\inn}+\frac{1}{2}\frac{d^{2}}{dX^{2}}T_{\inn}+\left(\frac{d}{dX}u_{\inn}\right)^{2}=0.
  \end{equation} 
\end{subequations}
We have to solve the above equations with the boundary conditions~\eqref{eq:evol-bc-a} at
$X=0$, that is,
\begin{equation}
  u_{\inn}(X=0)=0, \qquad T_{\inn}(X=0)=1,
\end{equation}
and the matching conditions \cite{bender_advanced_1999}
\begin{subequations}\label{eq:matching}
  \begin{equation}
  \lim_{X\to\infty}u_{\inn}(X)=\lim_{x\to
    0}u_{\out}(x)=\frac{\nu}{4}\sqrt{\pi T_{0}},
\end{equation}
\begin{equation}
  \lim_{X\to\infty}T_{\inn}(X)=\lim_{x\to 0}T_{\out}(x)=T_{0}.
\end{equation}
\end{subequations}
The latter conditions assure that the solution in the boundary layer
smoothly matches the outer solution in the bulk. These
matching conditions make it possible to determine the unknown
bulk temperature $T_{0}$, as shown below.

Equation~\eqref{eq:bl-eq-1} is integrated straightforwardly to give
\begin{equation}\label{eq:bl-eq-1-sol} \frac{d}{dX}u_{\inn}-2\sqrt{\frac{T_{\inn}}{\pi}}=-2\sqrt{\frac{T_{0}}{\pi}},
\end{equation}
where the constant on the rhs has been obtained by matching the outer
and inner solutions to the lowest order; note that
$du_{\inn}/dX \to 0$ for $X\to\infty$. Substitution of
\eqref{eq:bl-eq-1-sol} into \eqref{eq:bl-eq-2} yields, after some
simple algebra
\begin{equation}
  \frac{d^{2}}{dX^{2}}\theta+\frac{8}{\pi}
  \left( 1-\sqrt{\theta}\right)=0, \qquad \theta=\frac{T_{\inn}}{T_{0}}.
\end{equation}
A first integral can be directly derived from this equation,
\begin{equation} \frac{1}{2}\left(\frac{d}{dX}\theta\right)^{2}+\frac{8}{\pi}\left[\theta-\frac{2}{3}\theta^{3/2}\right]=D,
\end{equation}
where $D$ is a constant. Again, $D$ is obtained by matching arguments
in the limit as $X\to\infty$, for which we have that $\theta\to 1$ and
$d\theta/dX\to 0$. Therefore, $D=\frac{8}{3\pi}$ and
\begin{equation}\label{eq:bl-theta-X-eq}
\frac{d}{dX}\theta=-\frac{4}{\sqrt{\pi}}\sqrt{\frac{1}{3}+\frac{2}{3}\theta^{3/2}-\theta}.
\end{equation}
We have chosen the minus sign on the rhs because $d\theta/dX$ must be
negative in the boundary layer, since the bulk temperature $T_{0}<1$
as a consequence of the dissipative character of the dynamics.

We do not need to solve Eq.~\eqref{eq:bl-theta-X-eq} to obtain
$T_{0}$, which is our main goal. Going back to
Eq.~\eqref{eq:bl-eq-1-sol}, we can rewrite it as
\begin{equation} \frac{d}{dX}u_{\inn}=2\sqrt{\frac{T_{0}}{\pi}} \left(\sqrt{\theta}-1\right)
\end{equation}
and combining it with Eq.~\eqref{eq:bl-theta-X-eq},
\begin{equation}
  du_{\inn}=-\frac{\sqrt{T_{0}}}{2}\frac{\sqrt{\theta}-1}{\sqrt{\frac{1}{3}+\frac{2}{3}\theta^{3/2}-\theta}} d\theta.
\end{equation}
This equation allows us to calculate $u_{\inn}$ as a function of
$\theta$, taking into account that $u_{\inn}(X=0)=0$ and
$\theta(X=0)=T_{0}^{-1}$,
\begin{equation}
u_{\inn}(\theta)=\sqrt{T_{0}}\,\left[\,\sqrt{\frac{1}{3}+\frac{2}{3}T_{0}^{-3/2}-T_{0}^{-1}}-\sqrt{\frac{1}{3}+\frac{2}{3}\theta^{3/2}-\theta}\,\,\right].
\end{equation}
Now we impose the matching conditions in Eq.~\eqref{eq:matching}, that
is, $\{u_{\inn}\to \nu \sqrt{\pi T_{0}}/4,\theta\to 1\}$
in the limit as $X\to\infty$. Hence, one can write that
\begin{equation}\label{eq:T0-eq}
  \frac{\sqrt{\pi}}{4}\nu=\varphi(T_{0}), \qquad 
  \varphi(T_{0})=\sqrt{\frac{1}{3}+\frac{2}{3}T_{0}^{-3/2}-T_{0}^{-1}}.
\end{equation}
The function $\varphi(T_{0})$ on the rhs is a monotonically decreasing
function of $T_{0}$ (recall that $0\leq T_{0}\leq 1$),
\begin{equation}
  \lim_{T_{0}\to 0}\varphi(T_{0})=\infty, \quad
  \lim_{T_{0}\to 1}\varphi(T_{0})=0,
  \quad \frac{d\varphi(T_{0})}{dT_{0}}<0.
\end{equation}
Thus, Eq.~\eqref{eq:T0-eq} is the desired expression for the bulk
temperature as a function of the macroscopic inelasticity $\nu$, since
it univocally gives $T_{0}$ for each value of $\nu$.

In Fig.~\ref{t_prof2}, a good agreement is shown
between the prediction for $T_0$ obtained here and numerical
simulations. The discrepancies remain quite small for $\nu\lesssim 2$,
becoming only larger for the highly dissipative case $\nu=20$. It must
be taken into account that the bulk temperature $T_{0}$ has been
assumed to be of the order of unity in our theory, whereas
Eq.~\eqref{eq:T0-eq} implies that it becomes
very small for high $\nu$, specifically
\begin{equation}
  T_{0}\sim \left(\frac{32}{3\pi}\right)^{2/3} \nu^{-4/3}, \quad
  \nu\gg 1.
\end{equation}
This means that a more elaborate theory, corresponding to a different
dominant balance in the hydrodynamic equations, might be necessary in
the highly dissipative limit $\nu\gg 1$. See also next section for
other possible sources of discrepancy between our theory and the
numerical results as $\nu$ increase.

One can also examine the role of the boundary layers in
Fig.~\ref{t_prof2}.  The boundary layers are
barely noticeable for the smaller values of $\nu$, $\nu=0.2$ and
$\nu=0.02$, for which the bulk temperature is close to unity. In
addition, these small values of $\nu$ necessarily bring about a
stronger noise in the averages, since the bulk temperature is
larger. As $\nu$ increases, the bulk temperature decreases and so do
the fluctuations while the boundary layers become more visible. In
Figs.~\ref{fig:temp-prof} and~\ref{t_prof2}, it
seems that the boundary layer is wider for $\nu=2$ than for
$\nu=20$. This is reasonable and consistent with the behaviour observed
in other models in the strongly dissipative limit $\nu\gg 1$
\cite{prados_nonlinear_2012,hurtado_typical_2013}, in which the width
of the boundary layer algebraically decreases with $\nu$ and vanishes
in the limit as $\nu\to\infty$. Besides the above calculations show
that, as a function of the system size, the width of the boundary
layer in the $x$ variable is expected to be of order $L^{-1}$. This
scaling is not clearly confirmed in our simulations, which seem to
indicate that this width does not go to zero in the limit as
$L\to\infty$, especially for the highest macroscopic inelasticity
$\nu=20$. This discrepancy might also be mended by a more elaborate
theory in the highly dissipative case, see also next section.

Finally, it should also be noted some kind of ``boundary
resistance'', that is, a difference between the actual value of $T$ at
the boundary and the value imposed by the thermostat in the simulation
($T=1$). This phenomenon  is known to appear in nonlinear transport
problems~\cite{livi03,dhar_heat_2008} and has been already observed in
other models~\cite{prados_nonlinear_2012}.

\section{Non-Gaussian local distributions and spatial correlations.}\label{app:C}

\begin{figure}
\includegraphics[clip=true,width=3.25in]{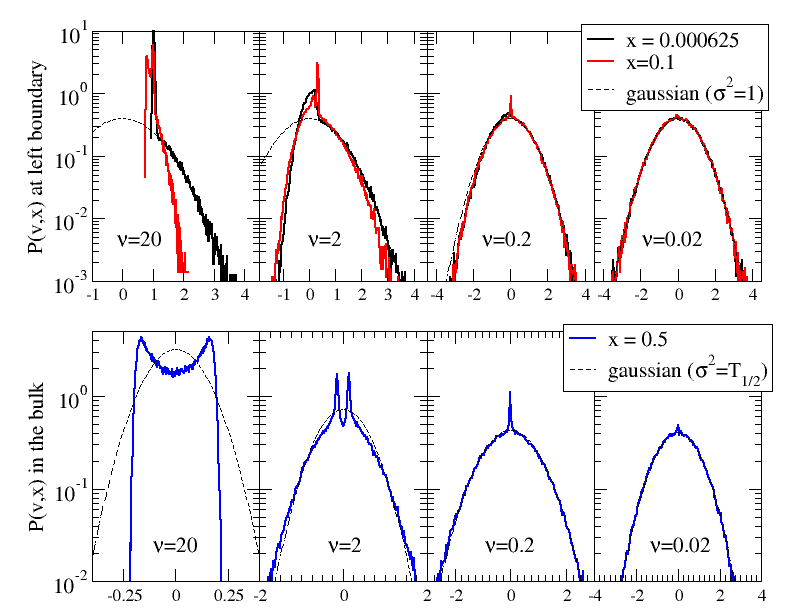}
\caption{\label{vel_pdf} Local velocity distributions $P(v,x)$ in the
  stationary state for different values of $\nu$ and different
  positions. All the plots corresponds to the largest considered
  system size $N=1600$. Top row: position close to the left boundary,
  $x\to 0$ and $x=0.1$. Bottom row: bulk position $x=0.5$. }
\end{figure}

Here we investigate the validity of the local equilibrium
approximation. A first check is obtained by measuring the local
velocity pdf. This is done in Fig.~\ref{vel_pdf}. In the top row, we
consider positions close to the boundaries and deviations from
Gaussianity are already apparent for $\nu=0.2$, with an anomalous but
localised peak at $v=0$. These discrepancies become even more patent
as $\nu$ increases, with the emergence of asymmetric tails in the
pdf. The pdf in the bulk of the system is presented in the bottom row
of the same figure. Again, discrepancies arise and are evident, with
an analogous localised peak at $v=0$ for $\nu=0.2$ that splits into
two symmetric peaks for larger $\nu\gtrsim
1$~\cite{mcnamara93,benedetto97}. Note that the tails remain
symmetric with respect to $v=0$ in the bulk, however.

Our conclusion is that some of the observed discrepancies as $\nu$
increases stem from the non-Gaussianities described above. Certainly,
the non-Gaussianities are more important at the boundary layers than
in the bulk, where the deviations seem to be milder, especially for
the not-so-large inelasticity $\nu=2$. Anyhow, non-Gaussianities may
be responsible for the deviations of the actual bulk temperature
$T_{1/2}$ from the theoretical description $T_{0}$ for $\nu\gtrsim
1$. It has to be taken into account that we have assumed that the
local equilibrium approximation holds both in the bulk and at the
boundary layers for deriving Eq.~\eqref{bulkt}. It is remarkable that
the relative error between $T_{0}$ and $T_{1/2}$ remains under ten per
cent for $\nu=2$, despite the large discrepancies at the boundaries
that include asymmetry with respect to $v=0$.

\begin{figure}
\includegraphics[clip=true,width=3.25in]{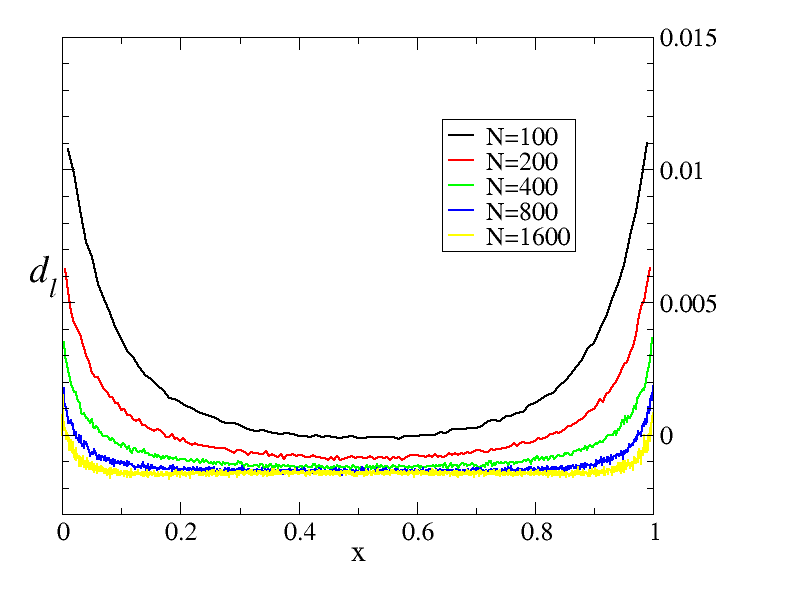}
\caption{\label{vel_corr} Nearest neighbour velocity correlation
 $d_{l}$ in the stationary state as a function of
 $x=l/L$, for different system sizes while keeping constant the
 macroscopic inelasticity $\nu=19$.  }
\end{figure}

Also, we have looked into the nearest-neighbour correlations
$d_{l}=\langle v_{l-1} v_{l} \rangle - \langle v_{l-1} \rangle \langle
v_{l} \rangle$, which have been assumed to vanish upon writing the
local equilibrium approximation \eqref{leq}. This assumption is
consistent with the Molecular Chaos hypothesis, in which these
correlations are assumed to be of the order of $L^{-1}$.  Taking into
account that we have incorporated $O(L^{-1})$ corrections into our
theory, these correlations are another possible source for
discrepancies and should be investigated.  In Fig.~\ref{vel_corr}, the
numerical evaluation of the nearest-neighbour correlations is
displayed. It is clearly seen that $d_{l}$ is always different from
zero and for large $L$ (keeping $\nu$ constant) displays a negative
plateau in the bulk. On the one hand, the value of the correlations at
such a plateau seems to be independent of $L$, which means that
Molecular Chaos is violated in the bulk. On the other hand, the
measured value is rather small and thus it seems that correlations are
not the main source for the observed discrepancies.

\bibliography{biblio,Mi-biblioteca-20-mar-2018}

\begin{thebibliography}{90}%
\makeatletter
\providecommand \@ifxundefined [1]{%
 \@ifx{#1\undefined}
}%
\providecommand \@ifnum [1]{%
 \ifnum #1\expandafter \@firstoftwo
 \else \expandafter \@secondoftwo
 \fi
}%
\providecommand \@ifx [1]{%
 \ifx #1\expandafter \@firstoftwo
 \else \expandafter \@secondoftwo
 \fi
}%
\providecommand \natexlab [1]{#1}%
\providecommand \enquote  [1]{``#1''}%
\providecommand \bibnamefont  [1]{#1}%
\providecommand \bibfnamefont [1]{#1}%
\providecommand \citenamefont [1]{#1}%
\providecommand \href@noop [0]{\@secondoftwo}%
\providecommand \href [0]{\begingroup \@sanitize@url \@href}%
\providecommand \@href[1]{\@@startlink{#1}\@@href}%
\providecommand \@@href[1]{\endgroup#1\@@endlink}%
\providecommand \@sanitize@url [0]{\catcode `\\12\catcode `\$12\catcode
  `\&12\catcode `\#12\catcode `\^12\catcode `\_12\catcode `\%12\relax}%
\providecommand \@@startlink[1]{}%
\providecommand \@@endlink[0]{}%
\providecommand \url  [0]{\begingroup\@sanitize@url \@url }%
\providecommand \@url [1]{\endgroup\@href {#1}{\urlprefix }}%
\providecommand \urlprefix  [0]{URL }%
\providecommand \Eprint [0]{\href }%
\providecommand \doibase [0]{http://dx.doi.org/}%
\providecommand \selectlanguage [0]{\@gobble}%
\providecommand \bibinfo  [0]{\@secondoftwo}%
\providecommand \bibfield  [0]{\@secondoftwo}%
\providecommand \translation [1]{[#1]}%
\providecommand \BibitemOpen [0]{}%
\providecommand \bibitemStop [0]{}%
\providecommand \bibitemNoStop [0]{.\EOS\space}%
\providecommand \EOS [0]{\spacefactor3000\relax}%
\providecommand \BibitemShut  [1]{\csname bibitem#1\endcsname}%
\let\auto@bib@innerbib\@empty
\bibitem [{\citenamefont {Huang}(1988)}]{huang}%
  \BibitemOpen
  \bibfield  {author} {\bibinfo {author} {\bibfnamefont {K.}~\bibnamefont
  {Huang}},\ }\href@noop {} {\emph {\bibinfo {title} {Statistical Mechanics}}}\
  (\bibinfo  {publisher} {John Wiley \& Sons},\ \bibinfo {year}
  {1988})\BibitemShut {NoStop}%
\bibitem [{\citenamefont {McLennan}(1989)}]{m89}%
  \BibitemOpen
  \bibfield  {author} {\bibinfo {author} {\bibfnamefont {J.~A.}\ \bibnamefont
  {McLennan}},\ }\href@noop {} {\emph {\bibinfo {title} {Introduction to
  Nonequilibrium Statistical Mechanics}}}\ (\bibinfo  {publisher}
  {Prentice-Hall},\ \bibinfo {year} {1989})\BibitemShut {NoStop}%
\bibitem [{\citenamefont {Jaeger}\ \emph {et~al.}(1996)\citenamefont {Jaeger},
  \citenamefont {Nagel},\ and\ \citenamefont {Behringer}}]{jaeger96b}%
  \BibitemOpen
  \bibfield  {author} {\bibinfo {author} {\bibfnamefont {H.~M.}\ \bibnamefont
  {Jaeger}}, \bibinfo {author} {\bibfnamefont {S.~R.}\ \bibnamefont {Nagel}}, \
  and\ \bibinfo {author} {\bibfnamefont {R.~P.}\ \bibnamefont {Behringer}},\
  }\href@noop {} {\bibfield  {journal} {\bibinfo  {journal} {Rev. Mod. Phys.}\
  }\textbf {\bibinfo {volume} {68}},\ \bibinfo {pages} {1259} (\bibinfo {year}
  {1996})}\BibitemShut {NoStop}%
\bibitem [{\citenamefont {Brilliantov}\ and\ \citenamefont
  {P\"oschel}(2004)}]{poeschel}%
  \BibitemOpen
  \bibfield  {author} {\bibinfo {author} {\bibfnamefont {N.~V.}\ \bibnamefont
  {Brilliantov}}\ and\ \bibinfo {author} {\bibfnamefont {T.}~\bibnamefont
  {P\"oschel}},\ }\href@noop {} {\emph {\bibinfo {title} {Kinetic Theory of
  Granular Gases}}}\ (\bibinfo  {publisher} {Oxford University Press},\
  \bibinfo {year} {2004})\BibitemShut {NoStop}%
\bibitem [{\citenamefont {Puglisi}(2015)}]{puglio15}%
  \BibitemOpen
  \bibfield  {author} {\bibinfo {author} {\bibfnamefont {A.}~\bibnamefont
  {Puglisi}},\ }\href@noop {} {\emph {\bibinfo {title} {Transport and
  Fluctuations in Granular Fluids}}}\ (\bibinfo  {publisher}
  {Springer-Verlag},\ \bibinfo {year} {2015})\BibitemShut {NoStop}%
\bibitem [{\citenamefont {Dufty}(2001)}]{D01}%
  \BibitemOpen
  \bibfield  {author} {\bibinfo {author} {\bibfnamefont {J.~W.}\ \bibnamefont
  {Dufty}},\ }\href@noop {} {\bibfield  {journal} {\bibinfo  {journal} {Adv.
  Complex Sys.}\ }\textbf {\bibinfo {volume} {4}},\ \bibinfo {pages} {397}
  (\bibinfo {year} {2001})}\BibitemShut {NoStop}%
\bibitem [{\citenamefont {Goldhirsch}(2003)}]{goldhirsch03}%
  \BibitemOpen
  \bibfield  {author} {\bibinfo {author} {\bibfnamefont {I.}~\bibnamefont
  {Goldhirsch}},\ }\href@noop {} {\bibfield  {journal} {\bibinfo  {journal}
  {Annu. Rev. Fluid Mech.}\ }\textbf {\bibinfo {volume} {35}},\ \bibinfo
  {pages} {267} (\bibinfo {year} {2003})}\BibitemShut {NoStop}%
\bibitem [{\citenamefont {Brilliantov}\ and\ \citenamefont
  {Pöschel}(2004)}]{brilliantov_kinetic_2004}%
  \BibitemOpen
  \bibfield  {author} {\bibinfo {author} {\bibfnamefont {N.~V.}\ \bibnamefont
  {Brilliantov}}\ and\ \bibinfo {author} {\bibfnamefont {T.}~\bibnamefont
  {Pöschel}},\ }\href@noop {} {\emph {\bibinfo {title} {Kinetic {Theory} of
  {Granular} {Gases}}}}\ (\bibinfo  {publisher} {OUP Oxford},\ \bibinfo {year}
  {2004})\BibitemShut {NoStop}%
\bibitem [{\citenamefont {Brey}\ \emph {et~al.}(1997)\citenamefont {Brey},
  \citenamefont {Dufty},\ and\ \citenamefont {Santos}}]{brey_dissipative_1997}%
  \BibitemOpen
  \bibfield  {author} {\bibinfo {author} {\bibfnamefont {J.~J.}\ \bibnamefont
  {Brey}}, \bibinfo {author} {\bibfnamefont {J.~W.}\ \bibnamefont {Dufty}}, \
  and\ \bibinfo {author} {\bibfnamefont {A.}~\bibnamefont {Santos}},\ }\href
  {/citations?view_op=view_citation&continue=/scholar%3Fhl%3Den%26start%3D210%26as_sdt%3D0,5%26scilib%3D1&citilm=1&citation_for_view=fLRxLRkAAAAJ:9Nmd_mFXekcC&hl=en&oi=p}
  {\bibfield  {journal} {\bibinfo  {journal} {J. Stat. Phys.}\ }\textbf
  {\bibinfo {volume} {87}},\ \bibinfo {pages} {1051} (\bibinfo {year}
  {1997})}\BibitemShut {NoStop}%
\bibitem [{\citenamefont {Brey}\ \emph {et~al.}(1998)\citenamefont {Brey},
  \citenamefont {Dufty}, \citenamefont {Kim},\ and\ \citenamefont
  {Santos}}]{BDKS98}%
  \BibitemOpen
  \bibfield  {author} {\bibinfo {author} {\bibfnamefont {J.~J.}\ \bibnamefont
  {Brey}}, \bibinfo {author} {\bibfnamefont {J.~W.}\ \bibnamefont {Dufty}},
  \bibinfo {author} {\bibfnamefont {C.~S.}\ \bibnamefont {Kim}}, \ and\
  \bibinfo {author} {\bibfnamefont {A.}~\bibnamefont {Santos}},\ }\href@noop {}
  {\bibfield  {journal} {\bibinfo  {journal} {Phys. Rev. E}\ }\textbf {\bibinfo
  {volume} {58}},\ \bibinfo {pages} {4638} (\bibinfo {year}
  {1998})}\BibitemShut {NoStop}%
\bibitem [{\citenamefont {Dufty}\ and\ \citenamefont {Brey}(2002)}]{DB02}%
  \BibitemOpen
  \bibfield  {author} {\bibinfo {author} {\bibfnamefont {J.~W.}\ \bibnamefont
  {Dufty}}\ and\ \bibinfo {author} {\bibfnamefont {J.~J.}\ \bibnamefont
  {Brey}},\ }\href@noop {} {\bibfield  {journal} {\bibinfo  {journal} {J. Stat.
  Phys.}\ }\textbf {\bibinfo {volume} {109}},\ \bibinfo {pages} {433} (\bibinfo
  {year} {2002})}\BibitemShut {NoStop}%
\bibitem [{\citenamefont {Ernst}(1981)}]{ERNST19811}%
  \BibitemOpen
  \bibfield  {author} {\bibinfo {author} {\bibfnamefont {M.}~\bibnamefont
  {Ernst}},\ }\href {\doibase https://doi.org/10.1016/0370-1573(81)90002-8}
  {\bibfield  {journal} {\bibinfo  {journal} {Physics Reports}\ }\textbf
  {\bibinfo {volume} {78}},\ \bibinfo {pages} {1 } (\bibinfo {year}
  {1981})}\BibitemShut {NoStop}%
\bibitem [{\citenamefont {Ben-Naim}\ and\ \citenamefont
  {Krapivsky}(2000)}]{ben-naim_multiscaling_2000}%
  \BibitemOpen
  \bibfield  {author} {\bibinfo {author} {\bibfnamefont {E.}~\bibnamefont
  {Ben-Naim}}\ and\ \bibinfo {author} {\bibfnamefont {P.~L.}\ \bibnamefont
  {Krapivsky}},\ }\href {\doibase 10.1103/PhysRevE.61.R5} {\bibfield  {journal}
  {\bibinfo  {journal} {Physical Review E}\ }\textbf {\bibinfo {volume} {61}},\
  \bibinfo {pages} {R5} (\bibinfo {year} {2000})}\BibitemShut {NoStop}%
\bibitem [{\citenamefont {Ernst}\ and\ \citenamefont
  {Brito}(2002{\natexlab{a}})}]{EB02}%
  \BibitemOpen
  \bibfield  {author} {\bibinfo {author} {\bibfnamefont {M.~H.}\ \bibnamefont
  {Ernst}}\ and\ \bibinfo {author} {\bibfnamefont {R.}~\bibnamefont {Brito}},\
  }\href@noop {} {\bibfield  {journal} {\bibinfo  {journal} {Europhys. Lett.}\
  }\textbf {\bibinfo {volume} {58}},\ \bibinfo {pages} {182} (\bibinfo {year}
  {2002}{\natexlab{a}})}\BibitemShut {NoStop}%
\bibitem [{\citenamefont {Baldassarri}\ \emph
  {et~al.}(2002{\natexlab{a}})\citenamefont {Baldassarri}, \citenamefont
  {Marconi},\ and\ \citenamefont {Puglisi}}]{BMP02}%
  \BibitemOpen
  \bibfield  {author} {\bibinfo {author} {\bibfnamefont {A.}~\bibnamefont
  {Baldassarri}}, \bibinfo {author} {\bibfnamefont {U.~M.~B.}\ \bibnamefont
  {Marconi}}, \ and\ \bibinfo {author} {\bibfnamefont {A.}~\bibnamefont
  {Puglisi}},\ }\href@noop {} {\bibfield  {journal} {\bibinfo  {journal}
  {Europhys. Lett.}\ }\textbf {\bibinfo {volume} {58}},\ \bibinfo {pages} {14}
  (\bibinfo {year} {2002}{\natexlab{a}})}\BibitemShut {NoStop}%
\bibitem [{\citenamefont {Ben-Naim}\ and\ \citenamefont
  {Krapivsky}(2002)}]{ben-naim02}%
  \BibitemOpen
  \bibfield  {author} {\bibinfo {author} {\bibfnamefont {E.}~\bibnamefont
  {Ben-Naim}}\ and\ \bibinfo {author} {\bibfnamefont {P.}~\bibnamefont
  {Krapivsky}},\ }\href@noop {} {\bibfield  {journal} {\bibinfo  {journal}
  {Phys. Rev. E}\ }\textbf {\bibinfo {volume} {66}},\ \bibinfo {pages} {011309}
  (\bibinfo {year} {2002})}\BibitemShut {NoStop}%
\bibitem [{\citenamefont {Ernst}\ and\ \citenamefont
  {Brito}(2002{\natexlab{b}})}]{ernst_high-energy_2002}%
  \BibitemOpen
  \bibfield  {author} {\bibinfo {author} {\bibfnamefont {M.~H.}\ \bibnamefont
  {Ernst}}\ and\ \bibinfo {author} {\bibfnamefont {R.}~\bibnamefont {Brito}},\
  }\href {\doibase 10.1209/epl/i2002-00622-0} {\bibfield  {journal} {\bibinfo
  {journal} {EPL (Europhysics Letters)}\ }\textbf {\bibinfo {volume} {58}},\
  \bibinfo {pages} {182} (\bibinfo {year} {2002}{\natexlab{b}})}\BibitemShut
  {NoStop}%
\bibitem [{\citenamefont {Ernst}\ and\ \citenamefont
  {Brito}(2002{\natexlab{c}})}]{ernst_scaling_2002}%
  \BibitemOpen
  \bibfield  {author} {\bibinfo {author} {\bibfnamefont {M.~H.}\ \bibnamefont
  {Ernst}}\ and\ \bibinfo {author} {\bibfnamefont {R.}~\bibnamefont {Brito}},\
  }\href {\doibase 10.1023/A:1020437925931} {\bibfield  {journal} {\bibinfo
  {journal} {Journal of Statistical Physics}\ }\textbf {\bibinfo {volume}
  {109}},\ \bibinfo {pages} {407} (\bibinfo {year}
  {2002}{\natexlab{c}})}\BibitemShut {NoStop}%
\bibitem [{\citenamefont {Bobylev}\ and\ \citenamefont
  {Cercignani}(2003)}]{bobylev_self-similar_2003}%
  \BibitemOpen
  \bibfield  {author} {\bibinfo {author} {\bibfnamefont {A.~V.}\ \bibnamefont
  {Bobylev}}\ and\ \bibinfo {author} {\bibfnamefont {C.}~\bibnamefont
  {Cercignani}},\ }\href {\doibase 10.1023/A:1021031031038} {\bibfield
  {journal} {\bibinfo  {journal} {Journal of Statistical Physics}\ }\textbf
  {\bibinfo {volume} {110}},\ \bibinfo {pages} {333} (\bibinfo {year}
  {2003})}\BibitemShut {NoStop}%
\bibitem [{\citenamefont {Bobylev}\ \emph {et~al.}(2003)\citenamefont
  {Bobylev}, \citenamefont {Cercignani},\ and\ \citenamefont
  {Toscani}}]{bobylev_proof_2003}%
  \BibitemOpen
  \bibfield  {author} {\bibinfo {author} {\bibfnamefont {A.~V.}\ \bibnamefont
  {Bobylev}}, \bibinfo {author} {\bibfnamefont {C.}~\bibnamefont {Cercignani}},
  \ and\ \bibinfo {author} {\bibfnamefont {G.}~\bibnamefont {Toscani}},\ }\href
  {\doibase 10.1023/A:1022273528296} {\bibfield  {journal} {\bibinfo  {journal}
  {Journal of Statistical Physics}\ }\textbf {\bibinfo {volume} {111}},\
  \bibinfo {pages} {403} (\bibinfo {year} {2003})}\BibitemShut {NoStop}%
\bibitem [{\citenamefont {Bobylev}\ \emph {et~al.}(2009)\citenamefont
  {Bobylev}, \citenamefont {Cercignani},\ and\ \citenamefont
  {Gamba}}]{bobylev_self-similar_2009}%
  \BibitemOpen
  \bibfield  {author} {\bibinfo {author} {\bibfnamefont {A.~V.}\ \bibnamefont
  {Bobylev}}, \bibinfo {author} {\bibfnamefont {C.}~\bibnamefont {Cercignani}},
  \ and\ \bibinfo {author} {\bibfnamefont {I.~M.}\ \bibnamefont {Gamba}},\
  }\href {\doibase 10.1007/s00220-009-0876-3} {\bibfield  {journal} {\bibinfo
  {journal} {Communications in Mathematical Physics}\ }\textbf {\bibinfo
  {volume} {291}},\ \bibinfo {pages} {599} (\bibinfo {year}
  {2009})}\BibitemShut {NoStop}%
\bibitem [{\citenamefont {Ilyin}(2016)}]{ilyin_exact_2016}%
  \BibitemOpen
  \bibfield  {author} {\bibinfo {author} {\bibfnamefont {O.}~\bibnamefont
  {Ilyin}},\ }\href {\doibase 10.1007/s10955-016-1643-4} {\bibfield  {journal}
  {\bibinfo  {journal} {Journal of Statistical Physics}\ }\textbf {\bibinfo
  {volume} {165}},\ \bibinfo {pages} {755} (\bibinfo {year}
  {2016})}\BibitemShut {NoStop}%
\bibitem [{\citenamefont {Carrillo}\ \emph {et~al.}(2000)\citenamefont
  {Carrillo}, \citenamefont {Cercignani},\ and\ \citenamefont
  {Gamba}}]{carrillo_steady_2000}%
  \BibitemOpen
  \bibfield  {author} {\bibinfo {author} {\bibfnamefont {J.~A.}\ \bibnamefont
  {Carrillo}}, \bibinfo {author} {\bibfnamefont {C.}~\bibnamefont
  {Cercignani}}, \ and\ \bibinfo {author} {\bibfnamefont {I.~M.}\ \bibnamefont
  {Gamba}},\ }\href {\doibase 10.1103/PhysRevE.62.7700} {\bibfield  {journal}
  {\bibinfo  {journal} {Physical Review E}\ }\textbf {\bibinfo {volume} {62}},\
  \bibinfo {pages} {7700} (\bibinfo {year} {2000})}\BibitemShut {NoStop}%
\bibitem [{\citenamefont {Santos}\ and\ \citenamefont
  {Ernst}(2003)}]{santos_exact_2003}%
  \BibitemOpen
  \bibfield  {author} {\bibinfo {author} {\bibfnamefont {A.}~\bibnamefont
  {Santos}}\ and\ \bibinfo {author} {\bibfnamefont {M.~H.}\ \bibnamefont
  {Ernst}},\ }\href {\doibase 10.1103/PhysRevE.68.011305} {\bibfield  {journal}
  {\bibinfo  {journal} {Physical Review E}\ }\textbf {\bibinfo {volume} {68}},\
  \bibinfo {pages} {011305} (\bibinfo {year} {2003})}\BibitemShut {NoStop}%
\bibitem [{\citenamefont {Villani}(2006)}]{villani_mathematics_2006}%
  \BibitemOpen
  \bibfield  {author} {\bibinfo {author} {\bibfnamefont {C.}~\bibnamefont
  {Villani}},\ }\href {\doibase 10.1007/s10955-006-9038-6} {\bibfield
  {journal} {\bibinfo  {journal} {Journal of Statistical Physics}\ }\textbf
  {\bibinfo {volume} {124}},\ \bibinfo {pages} {781} (\bibinfo {year}
  {2006})}\BibitemShut {NoStop}%
\bibitem [{\citenamefont {Van~Noije}\ and\ \citenamefont
  {Ernst}(1998)}]{van_noije_velocity_1998}%
  \BibitemOpen
  \bibfield  {author} {\bibinfo {author} {\bibfnamefont {T.~P.~C.}\
  \bibnamefont {Van~Noije}}\ and\ \bibinfo {author} {\bibfnamefont {M.~H.}\
  \bibnamefont {Ernst}},\ }\href
  {/citations?view_op=view_citation&continue=/scholar%3Fhl%3Den%26start%3D210%26as_sdt%3D0,5%26scilib%3D1&citilm=1&citation_for_view=fLRxLRkAAAAJ:fEOibwPWpKIC&hl=en&oi=p}
  {\bibfield  {journal} {\bibinfo  {journal} {Granul. Matter}\ }\textbf
  {\bibinfo {volume} {1}},\ \bibinfo {pages} {57} (\bibinfo {year}
  {1998})}\BibitemShut {NoStop}%
\bibitem [{\citenamefont {Santos}\ and\ \citenamefont
  {Montanero}(2009)}]{SM09}%
  \BibitemOpen
  \bibfield  {author} {\bibinfo {author} {\bibfnamefont {A.}~\bibnamefont
  {Santos}}\ and\ \bibinfo {author} {\bibfnamefont {J.~M.}\ \bibnamefont
  {Montanero}},\ }\href@noop {} {\bibfield  {journal} {\bibinfo  {journal}
  {Gran. Matt.}\ }\textbf {\bibinfo {volume} {11}},\ \bibinfo {pages} {157}
  (\bibinfo {year} {2009})}\BibitemShut {NoStop}%
\bibitem [{\citenamefont {Goldhirsch}(1999)}]{G99}%
  \BibitemOpen
  \bibfield  {author} {\bibinfo {author} {\bibfnamefont {I.}~\bibnamefont
  {Goldhirsch}},\ }\href@noop {} {\bibfield  {journal} {\bibinfo  {journal}
  {Chaos}\ }\textbf {\bibinfo {volume} {9}},\ \bibinfo {pages} {659} (\bibinfo
  {year} {1999})}\BibitemShut {NoStop}%
\bibitem [{\citenamefont {Kadanoff}(1999)}]{K99}%
  \BibitemOpen
  \bibfield  {author} {\bibinfo {author} {\bibfnamefont {L.~P.}\ \bibnamefont
  {Kadanoff}},\ }\href@noop {} {\bibfield  {journal} {\bibinfo  {journal} {Rev.
  Mod. Phys.}\ }\textbf {\bibinfo {volume} {71}},\ \bibinfo {pages} {435}
  (\bibinfo {year} {1999})}\BibitemShut {NoStop}%
\bibitem [{\citenamefont {Serero}\ \emph {et~al.}(2006)\citenamefont {Serero},
  \citenamefont {Goldhirsch}, \citenamefont {Noskowicz},\ and\ \citenamefont
  {Tan}}]{serero_hydrodynamics_2006}%
  \BibitemOpen
  \bibfield  {author} {\bibinfo {author} {\bibfnamefont {D.}~\bibnamefont
  {Serero}}, \bibinfo {author} {\bibfnamefont {I.}~\bibnamefont {Goldhirsch}},
  \bibinfo {author} {\bibfnamefont {S.~H.}\ \bibnamefont {Noskowicz}}, \ and\
  \bibinfo {author} {\bibfnamefont {M.-L.}\ \bibnamefont {Tan}},\ }\href
  {\doibase 10.1017/S0022112006009281} {\bibfield  {journal} {\bibinfo
  {journal} {Journal of Fluid Mechanics}\ }\textbf {\bibinfo {volume} {554}},\
  \bibinfo {pages} {237} (\bibinfo {year} {2006})}\BibitemShut {NoStop}%
\bibitem [{\citenamefont {Argentina}\ \emph {et~al.}(2002)\citenamefont
  {Argentina}, \citenamefont {Clerc},\ and\ \citenamefont
  {Soto}}]{argentina02}%
  \BibitemOpen
  \bibfield  {author} {\bibinfo {author} {\bibfnamefont {M.}~\bibnamefont
  {Argentina}}, \bibinfo {author} {\bibfnamefont {M.~G.}\ \bibnamefont
  {Clerc}}, \ and\ \bibinfo {author} {\bibfnamefont {R.}~\bibnamefont {Soto}},\
  }\href@noop {} {\bibfield  {journal} {\bibinfo  {journal} {Phys. Rev. Lett.}\
  }\textbf {\bibinfo {volume} {89}},\ \bibinfo {pages} {044301} (\bibinfo
  {year} {2002})}\BibitemShut {NoStop}%
\bibitem [{\citenamefont {Eshuis}\ \emph {et~al.}(2010)\citenamefont {Eshuis},
  \citenamefont {van~der Meer}, \citenamefont {Alam}, \citenamefont {van
  Gerner}, \citenamefont {van~der Weele},\ and\ \citenamefont
  {Lohse}}]{lohse10}%
  \BibitemOpen
  \bibfield  {author} {\bibinfo {author} {\bibfnamefont {P.}~\bibnamefont
  {Eshuis}}, \bibinfo {author} {\bibfnamefont {D.}~\bibnamefont {van~der
  Meer}}, \bibinfo {author} {\bibfnamefont {M.}~\bibnamefont {Alam}}, \bibinfo
  {author} {\bibfnamefont {H.~J.}\ \bibnamefont {van Gerner}}, \bibinfo
  {author} {\bibfnamefont {K.}~\bibnamefont {van~der Weele}}, \ and\ \bibinfo
  {author} {\bibfnamefont {D.}~\bibnamefont {Lohse}},\ }\href@noop {}
  {\bibfield  {journal} {\bibinfo  {journal} {Phys. Rev. Lett.}\ }\textbf
  {\bibinfo {volume} {104}},\ \bibinfo {pages} {038001} (\bibinfo {year}
  {2010})}\BibitemShut {NoStop}%
\bibitem [{\citenamefont {Puglisi}\ \emph {et~al.}(2012)\citenamefont
  {Puglisi}, \citenamefont {Gnoli}, \citenamefont {Gradenigo}, \citenamefont
  {Sarracino},\ and\ \citenamefont {Villamaina}}]{puglisi12}%
  \BibitemOpen
  \bibfield  {author} {\bibinfo {author} {\bibfnamefont {A.}~\bibnamefont
  {Puglisi}}, \bibinfo {author} {\bibfnamefont {A.}~\bibnamefont {Gnoli}},
  \bibinfo {author} {\bibfnamefont {G.}~\bibnamefont {Gradenigo}}, \bibinfo
  {author} {\bibfnamefont {A.}~\bibnamefont {Sarracino}}, \ and\ \bibinfo
  {author} {\bibfnamefont {D.}~\bibnamefont {Villamaina}},\ }\href@noop {}
  {\bibfield  {journal} {\bibinfo  {journal} {J. Chem. Phys.}\ }\textbf
  {\bibinfo {volume} {014704}},\ \bibinfo {pages} {136} (\bibinfo {year}
  {2012})}\BibitemShut {NoStop}%
\bibitem [{\citenamefont {Baldassarri}\ \emph {et~al.}(2003)\citenamefont
  {Baldassarri}, \citenamefont {Marconi},\ and\ \citenamefont
  {Puglisi}}]{bald2003}%
  \BibitemOpen
  \bibfield  {author} {\bibinfo {author} {\bibfnamefont {A.}~\bibnamefont
  {Baldassarri}}, \bibinfo {author} {\bibfnamefont {U.~M.~B.}\ \bibnamefont
  {Marconi}}, \ and\ \bibinfo {author} {\bibfnamefont {A.}~\bibnamefont
  {Puglisi}},\ }in\ \href@noop {} {\emph {\bibinfo {booktitle} {Lecture Notes
  in Physics - Granular Gas Dynamics}}},\ Vol.\ \bibinfo {volume} {624}\
  (\bibinfo  {publisher} {Springer},\ \bibinfo {year} {2003})\BibitemShut
  {NoStop}%
\bibitem [{\citenamefont {Ben-Naim}\ \emph {et~al.}(1999)\citenamefont
  {Ben-Naim}, \citenamefont {Chen}, \citenamefont {Doolen},\ and\ \citenamefont
  {Redner}}]{ben-naim99}%
  \BibitemOpen
  \bibfield  {author} {\bibinfo {author} {\bibfnamefont {E.}~\bibnamefont
  {Ben-Naim}}, \bibinfo {author} {\bibfnamefont {S.~Y.}\ \bibnamefont {Chen}},
  \bibinfo {author} {\bibfnamefont {G.~D.}\ \bibnamefont {Doolen}}, \ and\
  \bibinfo {author} {\bibfnamefont {S.}~\bibnamefont {Redner}},\ }\href@noop {}
  {\bibfield  {journal} {\bibinfo  {journal} {Phys. Rev. Lett.}\ }\textbf
  {\bibinfo {volume} {83}},\ \bibinfo {pages} {4069} (\bibinfo {year}
  {1999})}\BibitemShut {NoStop}%
\bibitem [{\citenamefont {Ostojic}\ \emph {et~al.}(2004)\citenamefont
  {Ostojic}, \citenamefont {Panja},\ and\ \citenamefont
  {Nienhuis}}]{ostojic04}%
  \BibitemOpen
  \bibfield  {author} {\bibinfo {author} {\bibfnamefont {S.}~\bibnamefont
  {Ostojic}}, \bibinfo {author} {\bibfnamefont {D.}~\bibnamefont {Panja}}, \
  and\ \bibinfo {author} {\bibfnamefont {B.}~\bibnamefont {Nienhuis}},\
  }\href@noop {} {\bibfield  {journal} {\bibinfo  {journal} {Phys. Rev. E}\
  }\textbf {\bibinfo {volume} {69}},\ \bibinfo {pages} {041301} (\bibinfo
  {year} {2004})}\BibitemShut {NoStop}%
\bibitem [{\citenamefont {Dey}\ \emph {et~al.}(2011)\citenamefont {Dey},
  \citenamefont {Das},\ and\ \citenamefont {Rajesh}}]{dey_lattice_2011}%
  \BibitemOpen
  \bibfield  {author} {\bibinfo {author} {\bibfnamefont {S.}~\bibnamefont
  {Dey}}, \bibinfo {author} {\bibfnamefont {D.}~\bibnamefont {Das}}, \ and\
  \bibinfo {author} {\bibfnamefont {R.}~\bibnamefont {Rajesh}},\ }\href
  {\doibase 10.1209/0295-5075/93/44001} {\bibfield  {journal} {\bibinfo
  {journal} {EPL (Europhysics Letters)}\ }\textbf {\bibinfo {volume} {93}},\
  \bibinfo {pages} {44001} (\bibinfo {year} {2011})}\BibitemShut {NoStop}%
\bibitem [{\citenamefont {Baldassarri}\ \emph
  {et~al.}(2002{\natexlab{b}})\citenamefont {Baldassarri}, \citenamefont
  {Marconi},\ and\ \citenamefont {Puglisi}}]{BMP02b}%
  \BibitemOpen
  \bibfield  {author} {\bibinfo {author} {\bibfnamefont {A.}~\bibnamefont
  {Baldassarri}}, \bibinfo {author} {\bibfnamefont {U.~M.~B.}\ \bibnamefont
  {Marconi}}, \ and\ \bibinfo {author} {\bibfnamefont {A.}~\bibnamefont
  {Puglisi}},\ }\href@noop {} {\bibfield  {journal} {\bibinfo  {journal} {Phys.
  Rev. E}\ }\textbf {\bibinfo {volume} {65}},\ \bibinfo {pages} {051301}
  (\bibinfo {year} {2002}{\natexlab{b}})}\BibitemShut {NoStop}%
\bibitem [{\citenamefont {Astillero}\ and\ \citenamefont
  {Santos}(2012)}]{astillero_unsteady_2012}%
  \BibitemOpen
  \bibfield  {author} {\bibinfo {author} {\bibfnamefont {A.}~\bibnamefont
  {Astillero}}\ and\ \bibinfo {author} {\bibfnamefont {A.}~\bibnamefont
  {Santos}},\ }\href {\doibase 10.1103/PhysRevE.85.021302} {\bibfield
  {journal} {\bibinfo  {journal} {Physical Review E}\ }\textbf {\bibinfo
  {volume} {85}},\ \bibinfo {pages} {021302} (\bibinfo {year}
  {2012})}\BibitemShut {NoStop}%
\bibitem [{\citenamefont {Du}\ \emph {et~al.}(1995)\citenamefont {Du},
  \citenamefont {Li},\ and\ \citenamefont {Kadanoff}}]{du95}%
  \BibitemOpen
  \bibfield  {author} {\bibinfo {author} {\bibfnamefont {Y.}~\bibnamefont
  {Du}}, \bibinfo {author} {\bibfnamefont {H.}~\bibnamefont {Li}}, \ and\
  \bibinfo {author} {\bibfnamefont {L.~P.}\ \bibnamefont {Kadanoff}},\
  }\href@noop {} {\bibfield  {journal} {\bibinfo  {journal} {Phys. Rev. Lett.}\
  }\textbf {\bibinfo {volume} {74}},\ \bibinfo {pages} {1268} (\bibinfo {year}
  {1995})}\BibitemShut {NoStop}%
\bibitem [{\citenamefont {Puglisi}\ \emph {et~al.}(1998)\citenamefont
  {Puglisi}, \citenamefont {Loreto}, \citenamefont {Marconi}, \citenamefont
  {Petri},\ and\ \citenamefont {Vulpiani}}]{PLMPV98}%
  \BibitemOpen
  \bibfield  {author} {\bibinfo {author} {\bibfnamefont {A.}~\bibnamefont
  {Puglisi}}, \bibinfo {author} {\bibfnamefont {V.}~\bibnamefont {Loreto}},
  \bibinfo {author} {\bibfnamefont {U.~M.~B.}\ \bibnamefont {Marconi}},
  \bibinfo {author} {\bibfnamefont {A.}~\bibnamefont {Petri}}, \ and\ \bibinfo
  {author} {\bibfnamefont {A.}~\bibnamefont {Vulpiani}},\ }\href@noop {}
  {\bibfield  {journal} {\bibinfo  {journal} {Phys. Rev. Lett.}\ }\textbf
  {\bibinfo {volume} {81}},\ \bibinfo {pages} {3848} (\bibinfo {year}
  {1998})}\BibitemShut {NoStop}%
\bibitem [{\citenamefont {Puglisi}\ \emph {et~al.}(1999)\citenamefont
  {Puglisi}, \citenamefont {Loreto}, \citenamefont {Marconi},\ and\
  \citenamefont {Vulpiani}}]{puglisi99}%
  \BibitemOpen
  \bibfield  {author} {\bibinfo {author} {\bibfnamefont {A.}~\bibnamefont
  {Puglisi}}, \bibinfo {author} {\bibfnamefont {V.}~\bibnamefont {Loreto}},
  \bibinfo {author} {\bibfnamefont {U.~M.~B.}\ \bibnamefont {Marconi}}, \ and\
  \bibinfo {author} {\bibfnamefont {A.}~\bibnamefont {Vulpiani}},\ }\href@noop
  {} {\bibfield  {journal} {\bibinfo  {journal} {Phys. Rev. E}\ }\textbf
  {\bibinfo {volume} {59}},\ \bibinfo {pages} {5582} (\bibinfo {year}
  {1999})}\BibitemShut {NoStop}%
\bibitem [{Note1()}]{Note1}%
  \BibitemOpen
  \bibinfo {note} {Therein, a regularisation of the collision was adopted:
  impacts at very small relative velocities were considered elastic to avoid
  inelastic collapse~\cite {luding98f}.}\BibitemShut {Stop}%
\bibitem [{\citenamefont {Shinde}\ \emph {et~al.}(2007)\citenamefont {Shinde},
  \citenamefont {Das},\ and\ \citenamefont {Rajesh}}]{shinde_violation_2007}%
  \BibitemOpen
  \bibfield  {author} {\bibinfo {author} {\bibfnamefont {M.}~\bibnamefont
  {Shinde}}, \bibinfo {author} {\bibfnamefont {D.}~\bibnamefont {Das}}, \ and\
  \bibinfo {author} {\bibfnamefont {R.}~\bibnamefont {Rajesh}},\ }\href
  {\doibase 10.1103/PhysRevLett.99.234505} {\bibfield  {journal} {\bibinfo
  {journal} {Phys. Rev. Lett.}\ }\textbf {\bibinfo {volume} {99}},\ \bibinfo
  {pages} {234505} (\bibinfo {year} {2007})}\BibitemShut {NoStop}%
\bibitem [{\citenamefont {Shinde}\ \emph {et~al.}(2009)\citenamefont {Shinde},
  \citenamefont {Das},\ and\ \citenamefont {Rajesh}}]{shinde_equivalence_2009}%
  \BibitemOpen
  \bibfield  {author} {\bibinfo {author} {\bibfnamefont {M.}~\bibnamefont
  {Shinde}}, \bibinfo {author} {\bibfnamefont {D.}~\bibnamefont {Das}}, \ and\
  \bibinfo {author} {\bibfnamefont {R.}~\bibnamefont {Rajesh}},\ }\href
  {\doibase 10.1103/PhysRevE.79.021303} {\bibfield  {journal} {\bibinfo
  {journal} {Phys. Rev. E}\ }\textbf {\bibinfo {volume} {79}},\ \bibinfo
  {pages} {021303} (\bibinfo {year} {2009})}\BibitemShut {NoStop}%
\bibitem [{\citenamefont {Nie}\ \emph {et~al.}(2002)\citenamefont {Nie},
  \citenamefont {Ben-Naim},\ and\ \citenamefont {Chen}}]{nie02}%
  \BibitemOpen
  \bibfield  {author} {\bibinfo {author} {\bibfnamefont {X.}~\bibnamefont
  {Nie}}, \bibinfo {author} {\bibfnamefont {E.}~\bibnamefont {Ben-Naim}}, \
  and\ \bibinfo {author} {\bibfnamefont {S.}~\bibnamefont {Chen}},\ }\href@noop
  {} {\bibfield  {journal} {\bibinfo  {journal} {Phys. Rev. Lett.}\ }\textbf
  {\bibinfo {volume} {89}},\ \bibinfo {pages} {204301} (\bibinfo {year}
  {2002})}\BibitemShut {NoStop}%
\bibitem [{\citenamefont {Trizac}\ and\ \citenamefont
  {Barrat}(2000)}]{trizac00}%
  \BibitemOpen
  \bibfield  {author} {\bibinfo {author} {\bibfnamefont {E.}~\bibnamefont
  {Trizac}}\ and\ \bibinfo {author} {\bibfnamefont {A.}~\bibnamefont
  {Barrat}},\ }\href@noop {} {\bibfield  {journal} {\bibinfo  {journal}
  {Europhys. J. E}\ }\textbf {\bibinfo {volume} {3}},\ \bibinfo {pages} {291}
  (\bibinfo {year} {2000})}\BibitemShut {NoStop}%
\bibitem [{\citenamefont {Zaburdaev}\ \emph {et~al.}(2006)\citenamefont
  {Zaburdaev}, \citenamefont {Brinkmann},\ and\ \citenamefont
  {Herminghaus}}]{zaburdaev06}%
  \BibitemOpen
  \bibfield  {author} {\bibinfo {author} {\bibfnamefont {V.~Y.}\ \bibnamefont
  {Zaburdaev}}, \bibinfo {author} {\bibfnamefont {M.}~\bibnamefont
  {Brinkmann}}, \ and\ \bibinfo {author} {\bibfnamefont {S.}~\bibnamefont
  {Herminghaus}},\ }\href@noop {} {\bibfield  {journal} {\bibinfo  {journal}
  {Phys. Rev. Lett.}\ }\textbf {\bibinfo {volume} {97}},\ \bibinfo {pages}
  {018001} (\bibinfo {year} {2006})}\BibitemShut {NoStop}%
\bibitem [{\citenamefont {Nieuwstadt}\ and\ \citenamefont
  {Steketee}(1995)}]{nieuwstadt_selected_1995}%
  \BibitemOpen
  \bibinfo {editor} {\bibfnamefont {F.~T.}\ \bibnamefont {Nieuwstadt}}\ and\
  \bibinfo {editor} {\bibfnamefont {J.~A.}\ \bibnamefont {Steketee}},\ eds.,\
  \href {//www.springer.com/la/book/9780792332657} {\emph {\bibinfo {title}
  {Selected {Papers} of {J}. {M}. {Burgers}}}}\ (\bibinfo  {publisher}
  {Springer Netherlands},\ \bibinfo {year} {1995})\BibitemShut {NoStop}%
\bibitem [{\citenamefont {Su}\ and\ \citenamefont
  {Gardner}(1969)}]{su_kortewegvries_1969}%
  \BibitemOpen
  \bibfield  {author} {\bibinfo {author} {\bibfnamefont {C.~H.}\ \bibnamefont
  {Su}}\ and\ \bibinfo {author} {\bibfnamefont {C.~S.}\ \bibnamefont
  {Gardner}},\ }\href {\doibase 10.1063/1.1664873} {\bibfield  {journal}
  {\bibinfo  {journal} {Journal of Mathematical Physics}\ }\textbf {\bibinfo
  {volume} {10}},\ \bibinfo {pages} {536} (\bibinfo {year} {1969})}\BibitemShut
  {NoStop}%
\bibitem [{\citenamefont {Shandarin}\ and\ \citenamefont
  {Zeldovich}(1989)}]{shandarin_large-scale_1989}%
  \BibitemOpen
  \bibfield  {author} {\bibinfo {author} {\bibfnamefont {S.~F.}\ \bibnamefont
  {Shandarin}}\ and\ \bibinfo {author} {\bibfnamefont {Y.~B.}\ \bibnamefont
  {Zeldovich}},\ }\href {\doibase 10.1103/RevModPhys.61.185} {\bibfield
  {journal} {\bibinfo  {journal} {Reviews of Modern Physics}\ }\textbf
  {\bibinfo {volume} {61}},\ \bibinfo {pages} {185} (\bibinfo {year}
  {1989})}\BibitemShut {NoStop}%
\bibitem [{\citenamefont {Frachebourg}(1999)}]{frachebourg99}%
  \BibitemOpen
  \bibfield  {author} {\bibinfo {author} {\bibfnamefont {L.}~\bibnamefont
  {Frachebourg}},\ }\href@noop {} {\bibfield  {journal} {\bibinfo  {journal}
  {Phys. Rev. Lett.}\ }\textbf {\bibinfo {volume} {82}},\ \bibinfo {pages}
  {1502} (\bibinfo {year} {1999})}\BibitemShut {NoStop}%
\bibitem [{\citenamefont {Frachebourg}\ \emph {et~al.}(2000)\citenamefont
  {Frachebourg}, \citenamefont {Martin},\ and\ \citenamefont
  {Piasecki}}]{frachebourg00}%
  \BibitemOpen
  \bibfield  {author} {\bibinfo {author} {\bibfnamefont {L.}~\bibnamefont
  {Frachebourg}}, \bibinfo {author} {\bibfnamefont {P.}~\bibnamefont {Martin}},
  \ and\ \bibinfo {author} {\bibfnamefont {J.}~\bibnamefont {Piasecki}},\
  }\href@noop {} {\bibfield  {journal} {\bibinfo  {journal} {Phys. A.}\
  }\textbf {\bibinfo {volume} {279}},\ \bibinfo {pages} {69} (\bibinfo {year}
  {2000})}\BibitemShut {NoStop}%
\bibitem [{\citenamefont {Brey}\ \emph {et~al.}(2001)\citenamefont {Brey},
  \citenamefont {Ruiz-Montero},\ and\ \citenamefont {Moreno}}]{brey01d}%
  \BibitemOpen
  \bibfield  {author} {\bibinfo {author} {\bibfnamefont {J.~J.}\ \bibnamefont
  {Brey}}, \bibinfo {author} {\bibfnamefont {M.~J.}\ \bibnamefont
  {Ruiz-Montero}}, \ and\ \bibinfo {author} {\bibfnamefont {F.}~\bibnamefont
  {Moreno}},\ }\href@noop {} {\bibfield  {journal} {\bibinfo  {journal} {Phys.
  Rev. E}\ }\textbf {\bibinfo {volume} {63}},\ \bibinfo {pages} {061305}
  (\bibinfo {year} {2001})}\BibitemShut {NoStop}%
\bibitem [{\citenamefont {Efrati}\ \emph {et~al.}(2005)\citenamefont {Efrati},
  \citenamefont {Livne},\ and\ \citenamefont {Meerson}}]{meerson05b}%
  \BibitemOpen
  \bibfield  {author} {\bibinfo {author} {\bibfnamefont {E.}~\bibnamefont
  {Efrati}}, \bibinfo {author} {\bibfnamefont {E.}~\bibnamefont {Livne}}, \
  and\ \bibinfo {author} {\bibfnamefont {B.}~\bibnamefont {Meerson}},\
  }\href@noop {} {\bibfield  {journal} {\bibinfo  {journal} {Phys. Rev. Lett.}\
  }\textbf {\bibinfo {volume} {94}},\ \bibinfo {pages} {088001} (\bibinfo
  {year} {2005})}\BibitemShut {NoStop}%
\bibitem [{\citenamefont {Meerson}\ and\ \citenamefont
  {Puglisi}(2005)}]{meerson05}%
  \BibitemOpen
  \bibfield  {author} {\bibinfo {author} {\bibfnamefont {B.}~\bibnamefont
  {Meerson}}\ and\ \bibinfo {author} {\bibfnamefont {A.}~\bibnamefont
  {Puglisi}},\ }\href@noop {} {\bibfield  {journal} {\bibinfo  {journal}
  {Europhys. Lett.}\ }\textbf {\bibinfo {volume} {70}},\ \bibinfo {pages} {478}
  (\bibinfo {year} {2005})}\BibitemShut {NoStop}%
\bibitem [{\citenamefont {Fouxon}\ \emph {et~al.}(2007)\citenamefont {Fouxon},
  \citenamefont {Meerson}, \citenamefont {Assaf},\ and\ \citenamefont
  {Livne}}]{meerson07}%
  \BibitemOpen
  \bibfield  {author} {\bibinfo {author} {\bibfnamefont {I.}~\bibnamefont
  {Fouxon}}, \bibinfo {author} {\bibfnamefont {B.}~\bibnamefont {Meerson}},
  \bibinfo {author} {\bibfnamefont {M.}~\bibnamefont {Assaf}}, \ and\ \bibinfo
  {author} {\bibfnamefont {E.}~\bibnamefont {Livne}},\ }\href@noop {}
  {\bibfield  {journal} {\bibinfo  {journal} {Phys. Rev. E}\ }\textbf {\bibinfo
  {volume} {75}},\ \bibinfo {pages} {050301(R)} (\bibinfo {year}
  {2007})}\BibitemShut {NoStop}%
\bibitem [{\citenamefont {Puglisi}\ \emph {et~al.}(2008)\citenamefont
  {Puglisi}, \citenamefont {Assaf}, \citenamefont {Fouxon},\ and\ \citenamefont
  {Meerson}}]{meerson08}%
  \BibitemOpen
  \bibfield  {author} {\bibinfo {author} {\bibfnamefont {A.}~\bibnamefont
  {Puglisi}}, \bibinfo {author} {\bibfnamefont {M.}~\bibnamefont {Assaf}},
  \bibinfo {author} {\bibfnamefont {I.}~\bibnamefont {Fouxon}}, \ and\ \bibinfo
  {author} {\bibfnamefont {B.}~\bibnamefont {Meerson}},\ }\href@noop {}
  {\bibfield  {journal} {\bibinfo  {journal} {Phys. Rev. E}\ }\textbf {\bibinfo
  {volume} {77}},\ \bibinfo {pages} {021305} (\bibinfo {year}
  {2008})}\BibitemShut {NoStop}%
\bibitem [{\citenamefont {Prasad}\ \emph {et~al.}(2013)\citenamefont {Prasad},
  \citenamefont {Sabhapandit},\ and\ \citenamefont
  {Dhar}}]{prasad_high-energy_2013}%
  \BibitemOpen
  \bibfield  {author} {\bibinfo {author} {\bibfnamefont {V.~V.}\ \bibnamefont
  {Prasad}}, \bibinfo {author} {\bibfnamefont {S.}~\bibnamefont {Sabhapandit}},
  \ and\ \bibinfo {author} {\bibfnamefont {A.}~\bibnamefont {Dhar}},\ }\href
  {\doibase 10.1209/0295-5075/104/54003} {\bibfield  {journal} {\bibinfo
  {journal} {EPL}\ }\textbf {\bibinfo {volume} {104}},\ \bibinfo {pages}
  {54003} (\bibinfo {year} {2013})}\BibitemShut {NoStop}%
\bibitem [{\citenamefont {Prasad}\ \emph {et~al.}(2014)\citenamefont {Prasad},
  \citenamefont {Sabhapandit},\ and\ \citenamefont
  {Dhar}}]{prasad_driven_2014}%
  \BibitemOpen
  \bibfield  {author} {\bibinfo {author} {\bibfnamefont {V.~V.}\ \bibnamefont
  {Prasad}}, \bibinfo {author} {\bibfnamefont {S.}~\bibnamefont {Sabhapandit}},
  \ and\ \bibinfo {author} {\bibfnamefont {A.}~\bibnamefont {Dhar}},\ }\href
  {\doibase 10.1103/PhysRevE.90.062130} {\bibfield  {journal} {\bibinfo
  {journal} {Physical Review E}\ }\textbf {\bibinfo {volume} {90}},\ \bibinfo
  {pages} {062130} (\bibinfo {year} {2014})}\BibitemShut {NoStop}%
\bibitem [{\citenamefont {Lasanta}\ \emph
  {et~al.}(2015{\natexlab{a}})\citenamefont {Lasanta}, \citenamefont
  {Manacorda}, \citenamefont {Prados},\ and\ \citenamefont {Puglisi}}]{LMPP15}%
  \BibitemOpen
  \bibfield  {author} {\bibinfo {author} {\bibfnamefont {A.}~\bibnamefont
  {Lasanta}}, \bibinfo {author} {\bibfnamefont {A.}~\bibnamefont {Manacorda}},
  \bibinfo {author} {\bibfnamefont {A.}~\bibnamefont {Prados}}, \ and\ \bibinfo
  {author} {\bibfnamefont {A.}~\bibnamefont {Puglisi}},\ }\href@noop {}
  {\bibfield  {journal} {\bibinfo  {journal} {New J. Phys.}\ }\textbf {\bibinfo
  {volume} {17}},\ \bibinfo {pages} {083039} (\bibinfo {year}
  {2015}{\natexlab{a}})}\BibitemShut {NoStop}%
\bibitem [{\citenamefont {Manacorda}\ \emph
  {et~al.}(2016{\natexlab{a}})\citenamefont {Manacorda}, \citenamefont {Plata},
  \citenamefont {Lasanta}, \citenamefont {Puglisi},\ and\ \citenamefont
  {Prados}}]{MPLPP16}%
  \BibitemOpen
  \bibfield  {author} {\bibinfo {author} {\bibfnamefont {A.}~\bibnamefont
  {Manacorda}}, \bibinfo {author} {\bibfnamefont {C.~A.}\ \bibnamefont
  {Plata}}, \bibinfo {author} {\bibfnamefont {A.}~\bibnamefont {Lasanta}},
  \bibinfo {author} {\bibfnamefont {A.}~\bibnamefont {Puglisi}}, \ and\
  \bibinfo {author} {\bibfnamefont {A.}~\bibnamefont {Prados}},\ }\href@noop {}
  {\bibfield  {journal} {\bibinfo  {journal} {J. Stat. Phys.}\ }\textbf
  {\bibinfo {volume} {164}},\ \bibinfo {pages} {810} (\bibinfo {year}
  {2016}{\natexlab{a}})}\BibitemShut {NoStop}%
\bibitem [{\citenamefont {Plata}\ \emph {et~al.}(2016)\citenamefont {Plata},
  \citenamefont {Manacorda}, \citenamefont {Lasanta}, \citenamefont {Puglisi},\
  and\ \citenamefont {Prados}}]{PMLPP16}%
  \BibitemOpen
  \bibfield  {author} {\bibinfo {author} {\bibfnamefont {C.~A.}\ \bibnamefont
  {Plata}}, \bibinfo {author} {\bibfnamefont {A.}~\bibnamefont {Manacorda}},
  \bibinfo {author} {\bibfnamefont {A.}~\bibnamefont {Lasanta}}, \bibinfo
  {author} {\bibfnamefont {A.}~\bibnamefont {Puglisi}}, \ and\ \bibinfo
  {author} {\bibfnamefont {A.}~\bibnamefont {Prados}},\ }\href@noop {}
  {\bibfield  {journal} {\bibinfo  {journal} {J. Stat. Mech.}\ }\textbf
  {\bibinfo {volume} {2016}},\ \bibinfo {pages} {093203} (\bibinfo {year}
  {2016})}\BibitemShut {NoStop}%
\bibitem [{\citenamefont {Prasad}\ \emph
  {et~al.}(2017{\natexlab{a}})\citenamefont {Prasad}, \citenamefont
  {Sabhapandit}, \citenamefont {Dhar},\ and\ \citenamefont
  {Narayan}}]{prasad_driven_2016}%
  \BibitemOpen
  \bibfield  {author} {\bibinfo {author} {\bibfnamefont {V.~V.}\ \bibnamefont
  {Prasad}}, \bibinfo {author} {\bibfnamefont {S.}~\bibnamefont {Sabhapandit}},
  \bibinfo {author} {\bibfnamefont {A.}~\bibnamefont {Dhar}}, \ and\ \bibinfo
  {author} {\bibfnamefont {O.}~\bibnamefont {Narayan}},\ }\href
  {https://arxiv.org/abs/1606.09561} {\bibfield  {journal} {\bibinfo  {journal}
  {Phys. Rev. E}\ }\textbf {\bibinfo {volume} {95}},\ \bibinfo {pages} {022115}
  (\bibinfo {year} {2017}{\natexlab{a}})}\BibitemShut {NoStop}%
\bibitem [{\citenamefont {Prasad}\ \emph
  {et~al.}(2017{\natexlab{b}})\citenamefont {Prasad}, \citenamefont {Das},
  \citenamefont {Sabhapandit},\ and\ \citenamefont
  {Rajesh}}]{prasad_velocity_2017}%
  \BibitemOpen
  \bibfield  {author} {\bibinfo {author} {\bibfnamefont {V.~V.}\ \bibnamefont
  {Prasad}}, \bibinfo {author} {\bibfnamefont {D.}~\bibnamefont {Das}},
  \bibinfo {author} {\bibfnamefont {S.}~\bibnamefont {Sabhapandit}}, \ and\
  \bibinfo {author} {\bibfnamefont {R.}~\bibnamefont {Rajesh}},\ }\href
  {\doibase 10.1103/PhysRevE.95.032909} {\bibfield  {journal} {\bibinfo
  {journal} {Physical Review E}\ }\textbf {\bibinfo {volume} {95}},\ \bibinfo
  {pages} {032909} (\bibinfo {year} {2017}{\natexlab{b}})}\BibitemShut
  {NoStop}%
\bibitem [{\citenamefont {Plata}\ and\ \citenamefont
  {Prados}(2017{\natexlab{a}})}]{plata_global_2017}%
  \BibitemOpen
  \bibfield  {author} {\bibinfo {author} {\bibfnamefont {C.~A.}\ \bibnamefont
  {Plata}}\ and\ \bibinfo {author} {\bibfnamefont {A.}~\bibnamefont {Prados}},\
  }\href {\doibase 10.1103/PhysRevE.95.052121} {\bibfield  {journal} {\bibinfo
  {journal} {Physical Review E}\ }\textbf {\bibinfo {volume} {95}},\ \bibinfo
  {pages} {052121} (\bibinfo {year} {2017}{\natexlab{a}})}\BibitemShut
  {NoStop}%
\bibitem [{\citenamefont {Plata}\ and\ \citenamefont
  {Prados}(2017{\natexlab{b}})}]{plata_kovacs-like_2017}%
  \BibitemOpen
  \bibfield  {author} {\bibinfo {author} {\bibfnamefont {C.~A.}\ \bibnamefont
  {Plata}}\ and\ \bibinfo {author} {\bibfnamefont {A.}~\bibnamefont {Prados}},\
  }\href {\doibase 10.3390/e19100539} {\bibfield  {journal} {\bibinfo
  {journal} {Entropy}\ }\textbf {\bibinfo {volume} {19}},\ \bibinfo {pages}
  {539} (\bibinfo {year} {2017}{\natexlab{b}})}\BibitemShut {NoStop}%
\bibitem [{\citenamefont {Ben-Naim}\ \emph
  {et~al.}(2003{\natexlab{a}})\citenamefont {Ben-Naim}, \citenamefont
  {Krapivsky},\ and\ \citenamefont {Redner}}]{BENNAIM2003190}%
  \BibitemOpen
  \bibfield  {author} {\bibinfo {author} {\bibfnamefont {E.}~\bibnamefont
  {Ben-Naim}}, \bibinfo {author} {\bibfnamefont {P.}~\bibnamefont {Krapivsky}},
  \ and\ \bibinfo {author} {\bibfnamefont {S.}~\bibnamefont {Redner}},\ }\href
  {\doibase https://doi.org/10.1016/S0167-2789(03)00171-4} {\bibfield
  {journal} {\bibinfo  {journal} {Physica D: Nonlinear Phenomena}\ }\textbf
  {\bibinfo {volume} {183}},\ \bibinfo {pages} {190 } (\bibinfo {year}
  {2003}{\natexlab{a}})}\BibitemShut {NoStop}%
\bibitem [{\citenamefont {Ben-Naim}\ \emph
  {et~al.}(2003{\natexlab{b}})\citenamefont {Ben-Naim}, \citenamefont
  {Krapivsky}, \citenamefont {Vazquez},\ and\ \citenamefont
  {Redner}}]{BENNAIM200399}%
  \BibitemOpen
  \bibfield  {author} {\bibinfo {author} {\bibfnamefont {E.}~\bibnamefont
  {Ben-Naim}}, \bibinfo {author} {\bibfnamefont {P.}~\bibnamefont {Krapivsky}},
  \bibinfo {author} {\bibfnamefont {F.}~\bibnamefont {Vazquez}}, \ and\
  \bibinfo {author} {\bibfnamefont {S.}~\bibnamefont {Redner}},\ }\href
  {\doibase https://doi.org/10.1016/j.physa.2003.08.027} {\bibfield  {journal}
  {\bibinfo  {journal} {Physica A: Statistical Mechanics and its Applications}\
  }\textbf {\bibinfo {volume} {330}},\ \bibinfo {pages} {99 } (\bibinfo {year}
  {2003}{\natexlab{b}})}\BibitemShut {NoStop}%
\bibitem [{\citenamefont {Slanina}(2004)}]{slanina_inelastically_2004}%
  \BibitemOpen
  \bibfield  {author} {\bibinfo {author} {\bibfnamefont {F.}~\bibnamefont
  {Slanina}},\ }\href {\doibase 10.1103/PhysRevE.69.046102} {\bibfield
  {journal} {\bibinfo  {journal} {Physical Review E}\ }\textbf {\bibinfo
  {volume} {69}},\ \bibinfo {pages} {046102} (\bibinfo {year}
  {2004})}\BibitemShut {NoStop}%
\bibitem [{\citenamefont {Porfiri}\ \emph {et~al.}(2007)\citenamefont
  {Porfiri}, \citenamefont {Bollt},\ and\ \citenamefont
  {Stilwell}}]{porfiri_decline_2007}%
  \BibitemOpen
  \bibfield  {author} {\bibinfo {author} {\bibfnamefont {M.}~\bibnamefont
  {Porfiri}}, \bibinfo {author} {\bibfnamefont {E.~M.}\ \bibnamefont {Bollt}},
  \ and\ \bibinfo {author} {\bibfnamefont {D.~J.}\ \bibnamefont {Stilwell}},\
  }\href {\doibase 10.1140/epjb/e2007-00186-3} {\bibfield  {journal} {\bibinfo
  {journal} {The European Physical Journal B}\ }\textbf {\bibinfo {volume}
  {57}},\ \bibinfo {pages} {481} (\bibinfo {year} {2007})}\BibitemShut
  {NoStop}%
\bibitem [{\citenamefont {Török}\ \emph {et~al.}(2013)\citenamefont
  {Török}, \citenamefont {Iñiguez}, \citenamefont {Yasseri}, \citenamefont
  {San~Miguel}, \citenamefont {Kaski},\ and\ \citenamefont
  {Kertész}}]{torok_opinions_2013}%
  \BibitemOpen
  \bibfield  {author} {\bibinfo {author} {\bibfnamefont {J.}~\bibnamefont
  {Török}}, \bibinfo {author} {\bibfnamefont {G.}~\bibnamefont {Iñiguez}},
  \bibinfo {author} {\bibfnamefont {T.}~\bibnamefont {Yasseri}}, \bibinfo
  {author} {\bibfnamefont {M.}~\bibnamefont {San~Miguel}}, \bibinfo {author}
  {\bibfnamefont {K.}~\bibnamefont {Kaski}}, \ and\ \bibinfo {author}
  {\bibfnamefont {J.}~\bibnamefont {Kertész}},\ }\href {\doibase
  10.1103/PhysRevLett.110.088701} {\bibfield  {journal} {\bibinfo  {journal}
  {Physical Review Letters}\ }\textbf {\bibinfo {volume} {110}},\ \bibinfo
  {pages} {088701} (\bibinfo {year} {2013})}\BibitemShut {NoStop}%
\bibitem [{\citenamefont {I{\~{n}}iguez}\ \emph {et~al.}(2014)\citenamefont
  {I{\~{n}}iguez}, \citenamefont {T{\"o}r{\"o}k}, \citenamefont {Yasseri},
  \citenamefont {Kaski},\ and\ \citenamefont {Kert{\'e}sz}}]{Iniguez2014}%
  \BibitemOpen
  \bibfield  {author} {\bibinfo {author} {\bibfnamefont {G.}~\bibnamefont
  {I{\~{n}}iguez}}, \bibinfo {author} {\bibfnamefont {J.}~\bibnamefont
  {T{\"o}r{\"o}k}}, \bibinfo {author} {\bibfnamefont {T.}~\bibnamefont
  {Yasseri}}, \bibinfo {author} {\bibfnamefont {K.}~\bibnamefont {Kaski}}, \
  and\ \bibinfo {author} {\bibfnamefont {J.}~\bibnamefont {Kert{\'e}sz}},\
  }\href {\doibase 10.1140/epjds/s13688-014-0007-z} {\bibfield  {journal}
  {\bibinfo  {journal} {EPJ Data Science}\ }\textbf {\bibinfo {volume} {3}},\
  \bibinfo {pages} {7} (\bibinfo {year} {2014})}\BibitemShut {NoStop}%
\bibitem [{\citenamefont {Rozanova}(2012)}]{rozanova}%
  \BibitemOpen
  \bibfield  {author} {\bibinfo {author} {\bibfnamefont {O.}~\bibnamefont
  {Rozanova}},\ }\href@noop {} {\bibfield  {journal} {\bibinfo  {journal}
  {Nonlinearity}\ }\textbf {\bibinfo {volume} {25}},\ \bibinfo {pages} {1547}
  (\bibinfo {year} {2012})}\BibitemShut {NoStop}%
\bibitem [{Note2()}]{Note2}%
  \BibitemOpen
  \bibinfo {note} {By reason of symmetry or, alternatively, by imposing that
  total momentum vanishes, $\DOTSI \intop \ilimits@ _{0}^{1}dx\protect \tmspace
  +\thinmuskip {.1667em}u(x,t)=0$.}\BibitemShut {Stop}%
\bibitem [{Note3()}]{Note3}%
  \BibitemOpen
  \bibinfo {note} {From a mathematical point of view, this property stems from
  the fact that the hydrodynamic equations contain only first-order spatial
  derivatives, which make it impossible to fit all the boundary conditions at
  $x=0,L$ \cite {bender_advanced_1999}.}\BibitemShut {Stop}%
\bibitem [{Note4()}]{Note4}%
  \BibitemOpen
  \bibinfo {note} {The temperature of the thermostat only sets the scale of
  energy and therefore we are not losing any generality.}\BibitemShut {Stop}%
\bibitem [{\citenamefont {Bender}\ and\ \citenamefont
  {Orszag}(1999)}]{bender_advanced_1999}%
  \BibitemOpen
  \bibfield  {author} {\bibinfo {author} {\bibfnamefont {C.~M.}\ \bibnamefont
  {Bender}}\ and\ \bibinfo {author} {\bibfnamefont {S.~A.}\ \bibnamefont
  {Orszag}},\ }\href
  {/citations?view_op=view_citation&continue=/scholar%3Fhl%3Den%26start%3D10%26as_sdt%3D0,5%26scilib%3D1&citilm=1&citation_for_view=fLRxLRkAAAAJ:jU7OWUQzBzMC&hl=en&oi=p}
  {\emph {\bibinfo {title} {Advanced mathematical methods for scientists and
  engineers {I}: {Asymptotic} methods and perturbation theory}}}\ (\bibinfo
  {publisher} {Springer},\ \bibinfo {year} {1999})\BibitemShut {NoStop}%
\bibitem [{Note5()}]{Note5}%
  \BibitemOpen
  \bibinfo {note} {The width of the boundary layers typically scales as $\nu
  ^{-1/2}$, as shown for instance in Ref.~\cite {prados_nonlinear_2012} for the
  dissipative version of the Kipnis-Marchioro-Presutti model.}\BibitemShut
  {Stop}%
\bibitem [{\citenamefont {Prados}\ \emph {et~al.}(2012)\citenamefont {Prados},
  \citenamefont {Lasanta},\ and\ \citenamefont
  {Hurtado}}]{prados_nonlinear_2012}%
  \BibitemOpen
  \bibfield  {author} {\bibinfo {author} {\bibfnamefont {A.}~\bibnamefont
  {Prados}}, \bibinfo {author} {\bibfnamefont {A.}~\bibnamefont {Lasanta}}, \
  and\ \bibinfo {author} {\bibfnamefont {P.~I.}\ \bibnamefont {Hurtado}},\
  }\href
  {/citations?view_op=view_citation&continue=/scholar%3Fhl%3Den%26start%3D378%26as_sdt%3D0,5%26scilib%3D1&citilm=1&citation_for_view=fLRxLRkAAAAJ:M3ejUd6NZC8C&hl=en&oi=p}
  {\bibfield  {journal} {\bibinfo  {journal} {Phys. Rev. E}\ }\textbf {\bibinfo
  {volume} {86}},\ \bibinfo {pages} {031134} (\bibinfo {year}
  {2012})}\BibitemShut {NoStop}%
\bibitem [{Note6()}]{Note6}%
  \BibitemOpen
  \bibinfo {note} {In other lattice models, the typical scaling has been found
  to be $t\propto \epsilon ^3$ as in Refs.~\cite
  {prados_nonlinear_2012,lasanta_fluctuating_2015,manacorda_lattice_2016}.}\BibitemShut
  {Stop}%
\bibitem [{Note7()}]{Note7}%
  \BibitemOpen
  \bibinfo {note} {Again, this scaling for the macroscopic inelasticity is
  different from the one found in other models, as a consequence of the
  different scaling of the continuous time variable. Notwithstanding, the
  underlying microscopic dynamics is quasi-elastic in all cases.}\BibitemShut
  {Stop}%
\bibitem [{\citenamefont {Lasanta}\ \emph
  {et~al.}(2015{\natexlab{b}})\citenamefont {Lasanta}, \citenamefont
  {Manacorda}, \citenamefont {Prados},\ and\ \citenamefont
  {Puglisi}}]{lasanta_fluctuating_2015}%
  \BibitemOpen
  \bibfield  {author} {\bibinfo {author} {\bibfnamefont {A.}~\bibnamefont
  {Lasanta}}, \bibinfo {author} {\bibfnamefont {A.}~\bibnamefont {Manacorda}},
  \bibinfo {author} {\bibfnamefont {A.}~\bibnamefont {Prados}}, \ and\ \bibinfo
  {author} {\bibfnamefont {A.}~\bibnamefont {Puglisi}},\ }\href {\doibase
  10.1088/1367-2630/17/8/083039} {\bibfield  {journal} {\bibinfo  {journal}
  {New J. Phys.}\ }\textbf {\bibinfo {volume} {17}},\ \bibinfo {pages} {083039}
  (\bibinfo {year} {2015}{\natexlab{b}})}\BibitemShut {NoStop}%
\bibitem [{\citenamefont {Hurtado}\ \emph {et~al.}(2013)\citenamefont
  {Hurtado}, \citenamefont {Lasanta},\ and\ \citenamefont
  {Prados}}]{hurtado_typical_2013}%
  \BibitemOpen
  \bibfield  {author} {\bibinfo {author} {\bibfnamefont {P.~I.}\ \bibnamefont
  {Hurtado}}, \bibinfo {author} {\bibfnamefont {A.}~\bibnamefont {Lasanta}}, \
  and\ \bibinfo {author} {\bibfnamefont {A.}~\bibnamefont {Prados}},\ }\href
  {\doibase 10.1103/PhysRevE.88.022110} {\bibfield  {journal} {\bibinfo
  {journal} {Physical Review E}\ }\textbf {\bibinfo {volume} {88}},\ \bibinfo
  {pages} {022110} (\bibinfo {year} {2013})}\BibitemShut {NoStop}%
\bibitem [{\citenamefont {Livi}\ \emph {et~al.}(2003)\citenamefont {Livi},
  \citenamefont {Politi},\ and\ \citenamefont {Lepri}}]{livi03}%
  \BibitemOpen
  \bibfield  {author} {\bibinfo {author} {\bibfnamefont {R.}~\bibnamefont
  {Livi}}, \bibinfo {author} {\bibfnamefont {A.}~\bibnamefont {Politi}}, \ and\
  \bibinfo {author} {\bibfnamefont {S.}~\bibnamefont {Lepri}},\ }\href@noop {}
  {\bibfield  {journal} {\bibinfo  {journal} {Phys. Rep.}\ }\textbf {\bibinfo
  {volume} {377}},\ \bibinfo {pages} {1} (\bibinfo {year} {2003})}\BibitemShut
  {NoStop}%
\bibitem [{\citenamefont {Dhar}(2008)}]{dhar_heat_2008}%
  \BibitemOpen
  \bibfield  {author} {\bibinfo {author} {\bibfnamefont {A.}~\bibnamefont
  {Dhar}},\ }\href
  {/citations?view_op=view_citation&continue=/scholar%3Fhl%3Den%26start%3D170%26as_sdt%3D0,5%26scilib%3D1&citilm=1&citation_for_view=fLRxLRkAAAAJ:VaXvl8Fpj5cC&hl=en&oi=p}
  {\bibfield  {journal} {\bibinfo  {journal} {Advances in Physics}\ }\textbf
  {\bibinfo {volume} {57}},\ \bibinfo {pages} {457} (\bibinfo {year}
  {2008})}\BibitemShut {NoStop}%
\bibitem [{\citenamefont {McNamara}\ and\ \citenamefont
  {Young}(1993)}]{mcnamara93}%
  \BibitemOpen
  \bibfield  {author} {\bibinfo {author} {\bibfnamefont {S.}~\bibnamefont
  {McNamara}}\ and\ \bibinfo {author} {\bibfnamefont {W.~R.}\ \bibnamefont
  {Young}},\ }\href@noop {} {\bibfield  {journal} {\bibinfo  {journal} {Phys.
  Fluids A}\ }\textbf {\bibinfo {volume} {5}},\ \bibinfo {pages} {34} (\bibinfo
  {year} {1993})}\BibitemShut {NoStop}%
\bibitem [{\citenamefont {Benedetto}\ \emph {et~al.}(1997)\citenamefont
  {Benedetto}, \citenamefont {Caglioti},\ and\ \citenamefont
  {Pulvirenti}}]{benedetto97}%
  \BibitemOpen
  \bibfield  {author} {\bibinfo {author} {\bibfnamefont {D.}~\bibnamefont
  {Benedetto}}, \bibinfo {author} {\bibfnamefont {E.}~\bibnamefont {Caglioti}},
  \ and\ \bibinfo {author} {\bibfnamefont {M.}~\bibnamefont {Pulvirenti}},\
  }\href@noop {} {\bibfield  {journal} {\bibinfo  {journal} {Math. Mod. Num.
  Anal.}\ }\textbf {\bibinfo {volume} {31}},\ \bibinfo {pages} {615} (\bibinfo
  {year} {1997})}\BibitemShut {NoStop}%
\bibitem [{\citenamefont {Luding}\ and\ \citenamefont
  {McNamara}(1998)}]{luding98f}%
  \BibitemOpen
  \bibfield  {author} {\bibinfo {author} {\bibfnamefont {S.}~\bibnamefont
  {Luding}}\ and\ \bibinfo {author} {\bibfnamefont {S.}~\bibnamefont
  {McNamara}},\ }\href@noop {} {\bibfield  {journal} {\bibinfo  {journal}
  {Granular Matter}\ }\textbf {\bibinfo {volume} {1}},\ \bibinfo {pages} {113}
  (\bibinfo {year} {1998})},\ \bibinfo {note} {e-print
  cond-mat/9810009}\BibitemShut {NoStop}%
\bibitem [{\citenamefont {Manacorda}\ \emph
  {et~al.}(2016{\natexlab{b}})\citenamefont {Manacorda}, \citenamefont {Plata},
  \citenamefont {Lasanta}, \citenamefont {Puglisi},\ and\ \citenamefont
  {Prados}}]{manacorda_lattice_2016}%
  \BibitemOpen
  \bibfield  {author} {\bibinfo {author} {\bibfnamefont {A.}~\bibnamefont
  {Manacorda}}, \bibinfo {author} {\bibfnamefont {C.~A.}\ \bibnamefont
  {Plata}}, \bibinfo {author} {\bibfnamefont {A.}~\bibnamefont {Lasanta}},
  \bibinfo {author} {\bibfnamefont {A.}~\bibnamefont {Puglisi}}, \ and\
  \bibinfo {author} {\bibfnamefont {A.}~\bibnamefont {Prados}},\ }\href
  {\doibase 10.1007/s10955-016-1575-z} {\bibfield  {journal} {\bibinfo
  {journal} {J. Stat. Phys.}\ }\textbf {\bibinfo {volume} {164}},\ \bibinfo
  {pages} {810} (\bibinfo {year} {2016}{\natexlab{b}})}\BibitemShut {NoStop}%
\end{thebibliography}%

\end{document}